\documentclass[journal]{IEEEtran}

\usepackage{graphicx}
\usepackage{caption}
\usepackage{subcaption}
\usepackage{float}
\usepackage{physics}
\usepackage{cite}
\usepackage{amsmath,amsthm,amsfonts,amssymb,amscd}
\usepackage{xcolor}
\usepackage{bm}
\usepackage{cuted}
\usepackage{algorithm}
\usepackage[noend]{algpseudocode}
\usepackage{multirow}
\usepackage{verbatim}
\usepackage[hidelinks]{hyperref}
\usepackage[frak=esstix]{mathalpha}
\usepackage{soul}

\newtheorem{remark}{Remark}

\usepackage{mathtools}
\begin{document}
\title{A Constrained Formulation for Simultaneous Line Parameter Estimation and \textcolor{black}{Instrument Transformer} Calibration}

\author{Antos~Cheeramban~Varghese,~\IEEEmembership{Member,~IEEE,} Rajasekhar~Anguluri,~\IEEEmembership{Member,~IEEE,} and
        Anamitra~Pal,~\IEEEmembership{Senior~Member,~IEEE} 
        \vspace{-1em}
\thanks{This work was supported in part by the National Science Foundation
(NSF) Grant 
ECCS-2145063.

Antos Varghese (email: avarghe6@asu.edu) and Anamitra Pal (email: Anamitra.Pal@asu.edu) are with the School of Electrical, Computer, and Energy Engineering, Arizona State University, Tempe, AZ 85287, USA.

Rajasekhar Anguluri (email: rajangul@umbc.edu) is with the Department of Computer Science and Electrical Engineering at University of Maryland, Baltimore County, MD 21250, USA.} 
}

\markboth{Under Review}%
{Shell \MakeLowercase{\textit{et al.}}: Bare Demo of IEEEtran.cls for IEEE Journals}

\maketitle

\begin{abstract}
The process of calibrating instrument transformers (ITs) has been greatly simplified by using phasor measurement unit (PMU) 
data since this process eliminates the need for (a) additional hardware, and (b) taking ITs offline. \textcolor{black}{However, such simplification comes at the cost of knowing the line parameters, whose estimation using PMU data in turn requires calibrated ITs. 
To solve this interdependency problem, we propose a 
novel framework that incorporates power system domain knowledge as constraints to perform simultaneous line parameter estimation and IT calibration.}
We demonstrate the effectiveness of our approach with simulated and real PMU 
data as well as for a power system application that uses both PMU data and line parameter information.
\end{abstract}

\begin{IEEEkeywords}
  Calibration, 
  Instrument transformer, Line parameter estimation, Phasor measurement unit
\end{IEEEkeywords}

\IEEEpeerreviewmaketitle

\section{Introduction}
\label{Intro}

An integral component of the wide-area measurement system is the phasor measurement unit (PMU). 
The PMU receives downsampled voltage and current signals from voltage transformers (VTs) and current transformers (CTs), and produces time-synchronized voltage phasor, current phasor, frequency, and rate-of-change-of-frequency measurements.
VTs and CTs, collectively referred to as instrument transformers (ITs), are typically \textcolor{black}{installed in outdoor switchyards}, unlike the PMU which is placed in a temperature controlled environment.
Consequently, over-time, the actual IT scaling factors often deviate from their name-plate values \cite{phadke2017synchronized}. This deviation in the scaling of the ITs \textit{adversely affects every downstream application} that relies on PMU measurements.
Moreover, since the ITs are separate from the PMU, 
the IT \textit{scaling factor deviations} are
not bounded by the 
total vector error (TVE) that the IEEE/IEC 60255-118-1-2018
Standard \cite{IEEE_IEC2018PMU_Std} mandates.
The process of correcting for this non-ideal scaling of the ITs is known as \textit{IT calibration}.

\textcolor{black}{Calibration methodologies for ITs have evolved considerably over the past two decades. Early approaches relied on hard calibration, which required taking the IT offline and physically transporting it to a high-precision laboratory \cite{brandolini2009simple, crotti2017industrial, siegenthaler2017computer}. This was inefficient, time-consuming, and expensive, making it impractical for frequent or large-scale use.
To address these shortcomings, remote and online calibration techniques were developed \cite{pal2015online, shi2012adaptive, zhou2012calibrating, chatterjee2018error}, which used available field measurements to calibrate the ITs without taking them offline. However, these methods inherently assumed that the transmission line parameters were known, which is rarely the case in practice since line parameters vary with ambient conditions and conductor aging.
In parallel, a body of literature focused on estimating line parameters directly from PMU measurements \cite{wehenkel2020parameter, varghese2022transmission, gupta2021compound, sharma2024comparative}. However, \cite{wehenkel2020parameter, varghese2022transmission, gupta2021compound, sharma2024comparative}
assumed the ITs are either ideal or already well calibrated.}

\textcolor{black}{Recognizing the limitations of both of these approaches, more recent works have tried to solve these two interdependent problems - ``line parameter estimation" (LPE) and ``instrument transformer (IT) calibration" - jointly. 
Wu et al. \cite{wu2015simultaneous} were the first ones that attempted to solve this interdependent problem. 
They proposed a multi-step process that uses voltage and current phasors to carry out this joint estimation. However, the process developed by \cite{wu2015simultaneous} was highly sensitive to measurement noise and the authors of that paper left it for investigation in future research. 
Next, Khandeparkar et al. \cite{khandeparkar2016detection} proposed a \textit{bias error detection} (BED) test to identify the ITs that require calibration. Goklani et al. \cite{goklani2020instrument} showed that the BED test has limitations in case of voltage magnitudes and proposed a  \textit{midpoint voltage} (MPV) test for identifying ITs requiring calibration followed by joint LPE and IT calibration. 
However, \cite{khandeparkar2016detection, goklani2020instrument} assumed the from-end ITs to be error-free, which need not be the case in reality.
Wang et al. then proposed a joint framework for positive sequence LPE and IT calibration \cite{wang2019transmission}. Although their approach gave low errors for IT calibration, the errors in the line parameter estimates were comparatively large.
}

\textcolor{black}{Taking a slightly different approach, Gupta et al. proposed a method combining low rank approximation and data mining, and leveraging the redundancy among capacitive voltage transformers (CVTs) in a substation, to calibrate the CVTs \textit{without requiring knowledge of line parameters} \cite{gupta2021calibrating}. In a subsequent work, they extended this approach by using the calibrated CVTs to calibrate the current transformers (CTs) within the same substation \cite{gupta2023three}.
However, if a sufficiently accurate CVT set cannot be identified for a given substation, the method may yield unreliable CVT calibration results, which in turn can degrade the downstream CT calibration that depends on them.
Finally, a quantization-based method that accounted for errors in both from-end and to-end ITs for joint LPE and IT calibration was proposed by Varghese et al. in \cite{varghese2024LinearSLIC}. However, the additional errors that were introduced due to the quantization procedure lowered this method's accuracy.
These limitations motivate the need for a joint estimation framework that can simultaneously and accurately estimate both line parameters and IT correction factors using realistic, high-accuracy meters rather than requiring idealized, error-free instrumentation.
This paper addresses this need by:}
\textcolor{black}{
\begin{enumerate}
    \item Proposing a novel formulation for accurate simultaneous line parameter estimation and IT calibration (SLIC) that contains a regularization term which softly penalizes deviations from highly accurate yet imperfect revenue-quality meter (RQM)\footnote{\textcolor{black}{RQMs are high quality ITs that are typically present at critical points in the power system where ownership of electricity changes hands, or where precise billing and financial settlement are required.
    A likely location where RQMs exist in a modern transmission system are the ends of the tie-lines that connect one utility to another.}} readings, relaxing the idealistic perfect-meter assumption prevalent in prior literature.
    \item Employing domain knowledge-based physical relations as equality constraints inside the formulation to perform SLIC
    across the network using only a pair of RQMs.
    \item Demonstrating successful solution to the SLIC problem using both simulated and real PMU measurements.
\end{enumerate}
}

\textcolor{black}{We also highlight the positive impact of performing SLIC on the power system problem of linear state estimation (LSE). Specifically, we demonstrate how solving the SLIC problem for only the highest voltage network of the IEEE 118-bus system can improve the LSE results by more than 20\%.}

\section{Mathematical Modeling of the SLIC Problem}
\label{SLIC_modeling_section}
\begin{figure}[ht]
    \centering
    \includegraphics[width=0.485\textwidth]{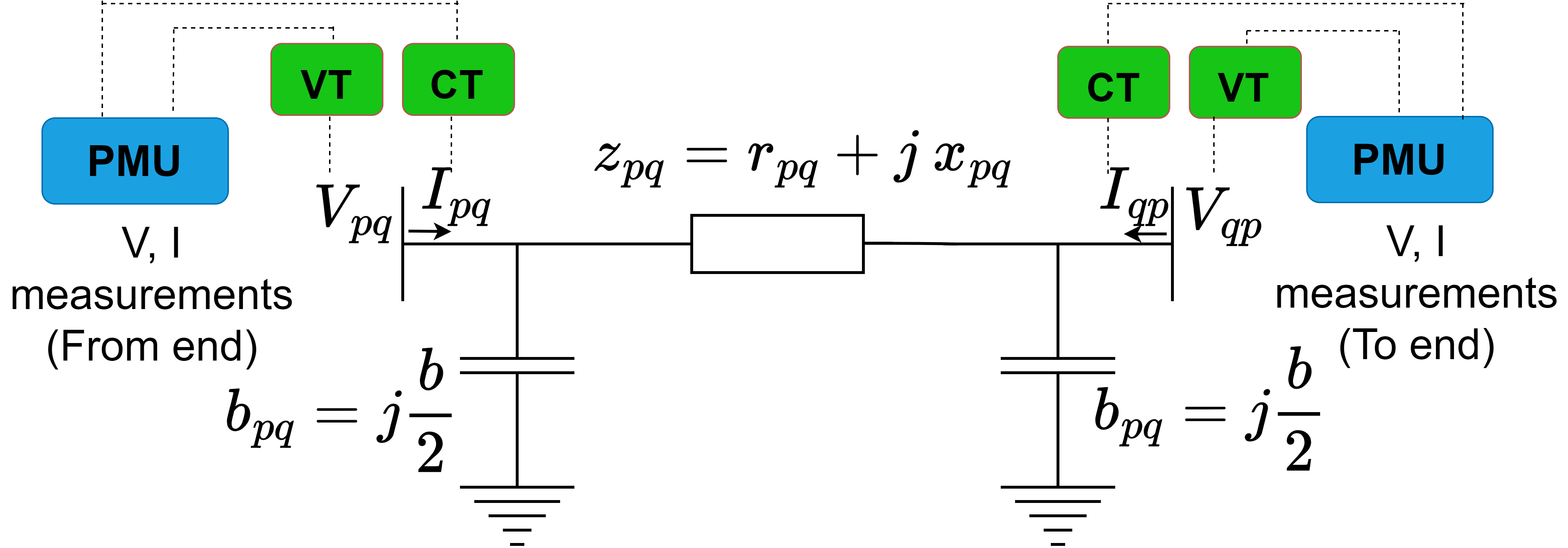}
    \caption{\textcolor{black}{$\pi$-model 
    of a transmission line used for explaining the proposed solution to the SLIC problem}} 
    \label{Medium_length_line_Segment_pi_with_IT}
    \vspace{-0.25em}
\end{figure}

Let us consider the medium length transmission line between buses $p$ and $q$
as shown in 
Fig.~\ref{Medium_length_line_Segment_pi_with_IT}. The line has PMUs (in blue color) on either end, while the ITs associated with the PMUs are shown in green.
The unknown parameters of this $p$-$q$ line/branch are resistance ($r_{pq}$), reactance ($x_{pq}$), and susceptance ($b_{pq}$).
By applying Kirchhoff's laws at both ends of the line, the $p$-th and $q$-th end \textit{true} current phasors ($I_{pq}^*$ and $I_{qp}^*$) 
can be written in terms of the corresponding \textit{true} voltage phasors ($V_{pq}^*$ and $V_{qp}^*$) and the line parameters as:

\begin{equation}
\label{eqn:SLIC_Basic_Ideal_Eqs}
         \begin{aligned}
         I_{pq}^* &= b_{pq} V_{pq}^* + (V_{pq}^* - V_{qp}^*)/z_{pq}\\
         I_{qp}^* &= b_{pq} V_{qp}^* - (V_{pq}^* - V_{qp}^*)/z_{pq}, 
         \end{aligned}
\end{equation}
where $z_{pq}=r_{pq}+jx_{pq}$,
and the superscript $^*$ indicates non-erroneous measurements obtained from perfect ITs and in the absence of any
measurement noise.

In reality,
IT scaling factors often differ from their name-plate values due to aging and environmental conditions.
These deviations in the IT scaling factors, termed \textit{ratio error} and denoted by $\eta$ in this paper, appear as an unknown \textit{multiplier} with the true phasor \cite{wang2019transmission}.
The allowable range in which $\eta$ can vary is specified in 
the IEEE C57.13.2016 Standard  
\cite{IEEE_C57_13_2016_std_for_ITs}.
Similarly, the noise coming from the PMU,
denoted by $\Delta$ in this paper, is \textit{additive} in nature \cite{frigo2024combining}. The net effect is that the measured phasor at the output of the PMU
has a \textit{composite} noise model: the multiplicative component $\eta$ due to the ITs and the additive component $\Delta$ due to the PMU.
As an example,
the measured phasor, $V_{pq}$, can be expressed in terms of the true phasor, $V_{pq}^*$, 
as:
$V_{pq} = \eta_{V_{pq}} V_{pq}^* + \Delta V_{{pq}}$.

\textcolor{black}{Note that $\eta$ remains constant over long time periods (e.g., weeks), but $\Delta$ changes with every PMU measurement \cite{varghese2024LinearSLIC}.
This occurs because the drifting of the IT scaling factors from their ideal values occurs over a very slow timescale (e.g., weeks to months), whereas the PMUs produce measurements at a much higher rate.}
Thus, to calibrate ITs, we must find
appropriate \textit{constant correction factors} from consecutive PMU measurements.
The correction factors, $\alpha$ for voltage and $\beta$ for currents, are the inverse of $\eta$. That is, $\alpha_{pq}=1/{\eta_{V_{pq}}}$ and $\beta_{pq}=1/{\eta_{I_{pq}}}$,
and $V_{pq}^* = \alpha_{pq}  (V_{pq}  - \Delta V_{{pq}})$ and $I_{pq}^* = \beta_{pq}  (I_{pq}  - \Delta I_{{pq}})$.
Table \ref{Variable Summary Table} summarizes the variables and symbols used to denote the measurements and unknowns (line parameters and IT correction factors),
for the line $p$-$q$. 

\begin{table}[ht]
\caption{Variables used in the SLIC study for line $p$-$q$}
\vspace{-0.5em}
\begin{tabular}{|l|l|l|c|}
\hline
\multicolumn{1}{|l|}{}                                                                      & \multicolumn{1}{l|}{}                                                             & Variable                                                                  & Symbol \\ \hline
\multirow{4}{*}{\begin{tabular}[c]{@{}c@{}}Input PMU \\ measurements\end{tabular}}          & \multirow{2}{*}{\begin{tabular}[c]{@{}c@{}}Voltage \\ measurements\end{tabular}}  & From-end voltage                                                          & $V_{pq}$    \\ \cline{3-4} 
                                                                                            &                                                                                   & To-end voltage                                                            & $V_{qp}$    \\ \cline{2-4} 
                                                                                            & \multirow{2}{*}{\begin{tabular}[c]{@{}c@{}}Current \\ measurements\end{tabular}}  & From-end current                                                          & $I_{pq}$    \\ \cline{3-4} 
                                                                                            &                                                                                   & To-end current                                                            & $I_{qp}$    \\ \hline
\multirow{7}{*}{\begin{tabular}[c]{@{}c@{}}SLIC parameters\\  to be estimated\end{tabular}} & \multirow{3}{*}{Line parameters}                                                  & Resistance                                                                & $r_{pq}$      \\ \cline{3-4} 
                                                                                            &                                                                                   & Reactance                                                                 & $x_{pq}$      \\ \cline{3-4} 
                                                                                            &                                                                                   & Shunt susceptance                                                         & $b_{pq}$      \\ \cline{2-4} 
                                                                                            & \multirow{4}{*}{\begin{tabular}[c]{@{}c@{}}IT correction\\  factors (CFs) \end{tabular}} & From-end VT CFs & $\alpha_{pq}$  \\ \cline{3-4} 
                                                                                            &                                                                                   & To-end VT CFs                                                            & $\alpha_{qp}$   \\ \cline{3-4} 
                                                                                            &                                                                                   & From-end CT CFs                                                          & $\beta_{pq}$  \\ \cline{3-4} 
                                                                                            &                                                                                   & To-end CT CFs                                                            & $\beta_{qp}$\\ \hline
\end{tabular}
\label{Variable Summary Table}
\end{table}

Now, replacing the non-erroneous measurements in \eqref{eqn:SLIC_Basic_Ideal_Eqs} with their composite noise-embedded counterparts, we get:
\begin{equation}
\label{eqn:SLIC_Basic_Realistic_Eqs}
         \begin{aligned}
         \beta_{pq} (I_{pq} -\Delta I_{pq}) &= b_{pq}\alpha_{pq} (V_{pq} - \Delta V_{pq}) + \big( \alpha_{pq} (V_{pq} -\Delta V_{pq}) \\
         &\quad - \alpha_{qp} (V_{qp} - \Delta V_{qp}) \big) /z_{pq}\\
         \beta_{qp} (I_{qp} - \Delta I_{qp}) &= b_{pq} \alpha_{qp} (V_{qp} - \Delta V_{qp}) - \big( \alpha_{pq} (V_{pq} - \Delta V_{pq})  \\
         &\quad - \alpha_{qp} (V_{qp} - \Delta V_{qp}) \big)/z_{pq}.
         \end{aligned}
\end{equation}

Multiplying the first sub-equation in \eqref{eqn:SLIC_Basic_Realistic_Eqs} by 
$z_{pq} (1+z_{pq} b_{pq})$ and the second sub-equation in \eqref{eqn:SLIC_Basic_Realistic_Eqs} by $z_{pq}$, and rearranging the terms, 
we get:
\begin{equation}
\label{eqn:SLIC_modeling_101_Eq_2}
         \begin{aligned}
           &(1+z_{pq} b_{pq})^2  \alpha_{pq} (V_{pq}  - \Delta V_{pq}) - (1+z_{pq} b_{pq}) \alpha_{qp} (V_{qp} - \Delta V_{qp}) \\&-  z_{pq} (1+z_{pq} b_{pq}) \beta_{pq} (I_{pq} - \Delta I_{{pq}}) = 0\\
           &(1+z_{pq} b_{pq}) \alpha_{qp} (V_{qp} - \Delta V_{qp}) - z_{pq}  \beta_{qp} (I_{qp} - \Delta I_{{qp}}) \\&= \alpha_{pq} (V_{pq}  - \Delta V_{pq}).
         \end{aligned}
\end{equation}

Now, \eqref{eqn:SLIC_modeling_101_Eq_2} cannot be used to solve for the SLIC parameters 
because the IT correction factors 
are present in all the terms.
We overcome this issue by assuming that an RQM-VT and an RQM-CT are already placed at the $p$-end of the line. This makes it possible to estimate the \textit{correction factor ratios} (CFRs) of all the ITs w.r.t. the RQM-VT.
Note that since RQMs have very small errors, 
the average value of the CFRs computed over multiple runs become
a good approximation for the correction factors themselves.
This strategy also takes advantage of
the fact that changes in $\eta$ occur at orders of magnitude lower rate than the speed with which a PMU produces its outputs.
In accordance with this strategy, every term in \eqref{eqn:SLIC_modeling_101_Eq_2} is divided by $\alpha_{pq} \ne 0$
to obtain:
\begin{equation}
\label{eqn:SLIC_modeling_101_Eq_3}
         \begin{aligned}
           &(1+z_{pq} b_{pq})^2  (V_{pq}  - \Delta V_{pq}) - (1+z_{pq} b_{pq}) \frac{\alpha_{qp}}{\alpha_{pq}} (V_{qp} - \Delta V_{qp})  \\& -z_{pq} (1+z_{pq} b_{pq}) \frac{\beta_{pq}}{\alpha_{pq}}  (I_{pq} - \Delta I_{{pq}})   = 0\\
           &(1+z_{pq} b_{pq}) \frac{\alpha_{qp}}{\alpha_{pq}} (V_{qp} - \Delta V_{qp}) - z_{pq}  \frac{\beta_{qp}}{\alpha_{pq}} (I_{qp} - \Delta I_{{qp}}) \\&=  (V_{pq}  - \Delta V_{pq}).
         \end{aligned}
\end{equation}

Eq. \eqref{eqn:SLIC_modeling_101_Eq_3} is in a form that can be solved to estimate
the CFRs of ITs w.r.t. the RQM-VT at bus $p$ as well as the series and shunt parameters of the line $p$-$q$.
However, apart from $b_{pq}$, all the other terms in \eqref{eqn:SLIC_modeling_101_Eq_3} are complex numbers.
Therefore, we separate the real and imaginary parts of \eqref{eqn:SLIC_modeling_101_Eq_3} and stack multiple instances of the resulting equation over time, to yield the following equation in matrix form:
\begin{equation}
\label{Abstracted_SLIC_linear}
\begin{aligned}
\begin{bmatrix} D_{pq} + \Delta D_{pq} \end{bmatrix} \begin{bmatrix}\theta_{pq} \end{bmatrix} = \begin{bmatrix}c_{pq} + \Delta c_{pq}\end{bmatrix}.
\end{aligned}
\end{equation}

The expressions for $D_{pq}$, $c_{pq}$, and $\theta_{pq}$ are shown in \eqref{D_in_SLIC_linear_form}\footnote{Due to space constraint, \eqref{D_in_SLIC_linear_form} is shown on the next page.}, \eqref{c_in_SLIC_linear_form}, and \eqref{theta_in_SLIC_linear_form}, respectively. 
In these equations, $n$ denotes the number of time-instants,
and $\Delta D_{pq}$ and $\Delta c_{pq}$ denote the additive noise in $D_{pq}$ and $c_{pq}$, respectively. 
\begin{figure*}[t]
\centering
\footnotesize{
\begin{equation}
\label{D_in_SLIC_linear_form}
\begin{aligned}
D_{pq} = 
 &\begin{bmatrix}      \mathcal{R}e( V_{pq}  (1)) & - \mathcal{I}m(V_{pq}  (1))  &  - \mathcal{R}e(V_{qp}  (1))  & \mathcal{I}m(V_{qp}  (1)) & -\mathcal{R}e(I_{pq}  (1)) & \mathcal{I}m(I_{pq}  (1)) & 0 & 0   \\ 0 & 0  & \mathcal{R}e(V_{qp}  (1)) &  -\mathcal{I}m(V_{qp}  (1)) & 0 & 0 & -\mathcal{R}e(I_{qp}  (1)) &\mathcal{I}m( I_{qp}  (1)) \\  \mathcal{I}m(V_{pq}  (1)) &  \mathcal{R}e(V_{pq}  (1)) &  - \mathcal{I}m(V_{qp}  (1)) & - \mathcal{R}e(V_{qp}  (1)) & - \mathcal{I}m(I_{pq}  (1)) & - \mathcal{R}e((I_{pq}  (1)) & 0 & 0 \\  0 & 0 & \mathcal{I}m(V_{qp}  (1)) & \mathcal{R}e((V_{qp}  (1)) & 0 & 0 & - \mathcal{I}m(I_{qp}  (1))  & - \mathcal{R}e(I_{qp}  (1))  \\
 \vdots & \vdots & \vdots &  \vdots & \vdots & \vdots & \vdots & \vdots \\
  \vdots & \vdots & \vdots &  \vdots & \vdots & \vdots & \vdots & \vdots \\
   \vdots & \vdots & \vdots &  \vdots & \vdots & \vdots & \vdots & \vdots \\
    \vdots & \vdots & \vdots &  \vdots & \vdots & \vdots & \vdots & \vdots \\
     \mathcal{R}e( V_{pq}  (n)) & - \mathcal{I}m(V_{pq}  (n))  &  - \mathcal{R}e(V_{qp}  (n))  & \mathcal{I}m(V_{qp}  (n)) & -\mathcal{R}e(I_{pq}  (n)) & \mathcal{I}m(I_{pq}  (n)) & 0 & 0   \\ 0 & 0  & \mathcal{R}e(V_{qp}  (n)) &  -\mathcal{I}m(V_{qp}  (n)) & 0 & 0 & -\mathcal{R}e(I_{qp}  (n)) &\mathcal{I}m( I_{qp}  (n)) \\  \mathcal{I}m(V_{pq}  (n)) &  \mathcal{R}e(V_{pq}  (n)) &  - \mathcal{I}m(V_{qp}  (n)) & - \mathcal{R}e(V_{qp}  (n)) & - \mathcal{I}m(I_{pq}  (n)) & - \mathcal{R}e((I_{pq}  (n)) & 0 & 0 \\  0 & 0 & \mathcal{I}m(V_{qp}  (n)) & \mathcal{R}e((V_{qp}  (n)) & 0 & 0 & - \mathcal{I}m(I_{qp}  (n))  & - \mathcal{R}e(I_{qp}  (n))  \end{bmatrix}. 
       \end{aligned}
\end{equation}}
\vspace{-2em}
\end{figure*}

\vspace{-1em}
\small
\begin{equation}
\label{c_in_SLIC_linear_form}
\begin{aligned}
c_{pq} ={\setlength{\arraycolsep}{3pt} \begin{bmatrix}  0   &     \mathcal{R}e( V_{pq} (1))& 0 &   \mathcal{I}m(V_{pq} (1))& \dots &   0   & \mathcal{I}m(V_{pq} (n)) \end{bmatrix}}
 \end{aligned}^T.
\end{equation}
\normalsize
\begin{equation}
\label{theta_in_SLIC_linear_form}
\begin{aligned}
    \theta_{pq} =   \begin{bmatrix} \theta_1 &  \theta_2 & \theta_3 & \theta_4 & \theta_5 & \theta_6 & \theta_7 & \theta_8 \end{bmatrix}  
 \end{aligned}^T.
\end{equation}

The relation between 
$\theta_{pq}$ and the unknown parameters (line parameters and correction factors) 
are specified below: 
\begin{equation}
\label{The_theta_explanation}
    \begin{aligned}
     \theta_1 +  j \theta_2 &=  (1+z_{pq} b_{pq})^2\\
      \theta_3 +  j \theta_4  &=  (1+z_{pq} b_{pq})  \frac{\alpha_{qp}}{\alpha_{pq}} \\
        \theta_5 +j  \theta_6  &=      z_{pq}(1+z_{pq} b_{pq})    \frac{\beta_{pq}}{\alpha_{pq}} \\
        \theta_7 +j  \theta_8 &=  z_{pq}    \frac{\beta_{qp}}{\alpha_{pq}}.
    \end{aligned}
\end{equation}

\section{Constrained Optimization Formulation for Solving SLIC Problem for RQM Branch}
\label{SLIC-RQM}
\textcolor{black}{The SLIC problem is solved in
two stages. In the first stage, the IT correction factor ratios are estimated along with the line parameters. In the second stage, the individual IT correction factors are recovered from these estimates.
Furthermore, note that \eqref{Abstracted_SLIC_linear} describes a linear system of equations whose solution results in \textit{eight ``knowns"} for the RQM branch $p$-$q$ (see \eqref{theta_in_SLIC_linear_form}).
Therefore, the goal of the first stage of SLIC}
is to estimate {\color{black}all \textit{nine ``unknowns"} of the branch:} line parameters ($r_{pq}, x_{pq},$ and $b_{pq}$) -- \textit{three} real numbers -- and  IT CFRs ($\frac{\alpha_{qp}}{\alpha_{pq}}$, $ \frac{\beta_{pq}}{\alpha_{pq}}$, and $ \frac{\beta_{qp}}{\alpha_{pq}}$) -- three complex numbers; that is, \textit{six} real numbers. 
Without additional information, usually in the form of some assumptions (such as presence of a ``perfect" IT), it is  difficult
to uniquely estimate all nine ``unknowns" for branch $p$-$q$ from the eight ``knowns" in \eqref{Abstracted_SLIC_linear}.
A strategy to circumvent this issue is described below. 

\vspace{-0.5em}
\subsection{Nonlinear Form for Solving the SLIC Problem}

We express $\theta_{pq}$ in terms of the actual physical parameters. Let $\psi$ be the vector of physical parameters for a transmission line. Then, $\psi_{pq}$ becomes:
\begin{equation}
\label{psi_Definition_9dim}
\begin{aligned}
\begin{bmatrix}
\psi_{pq}
\end{bmatrix} \triangleq 
\begin{bmatrix}
\psi_1 \\ \psi_2 \\ \psi_3 \\ \psi_4\\ \psi_5 \\ \psi_6\\ \psi_7 \\ \psi_8 \\ \psi_9
\end{bmatrix}
 = \begin{bmatrix}
r_{pq} \\ x_{pq} \\ b_{pq} \\ \mathcal{R}e \left(\frac{\alpha_{qp}}{\alpha_{pq}}\right) \\  \mathcal{I}m \left( \frac{\alpha_{qp}}{\alpha_{pq}}\right)  \\ \mathcal{R}e\left(  \frac{\beta_{pq}}{\alpha_{pq}}\right) \\ \mathcal{I}m\left(  \frac{\beta_{pq}}{\alpha_{pq}}\right) \\ \mathcal{R}e\left(  \frac{\beta_{qp}}{\alpha_{pq}}\right) \\ \mathcal{I}m\left(  \frac{\beta_{qp}}{\alpha_{pq}}\right) 
\end{bmatrix},
\end{aligned}
\end{equation}
where the first three variables in \eqref{psi_Definition_9dim} 
are the line parameters of the branch $p$-$q$, and the last six are the IT CFRs w.r.t. the VT at $p$-end of 
that branch.
The relation between $\theta$ and $\psi$ is: 
\begin{equation}
\begin{aligned}
\theta_{1}  &=  1 - 2 \psi_2 \psi_3 + \psi_2^2 \psi_3^2 - \psi_1^2 \psi_3^2\\
\theta_{2}  &= 2 \psi_1 \psi_3 - 2 \psi_1  \psi_2 \psi_3^2\\
\theta_{3}  &= \psi_4- \psi_2 \psi_3  \psi_4- \psi_1 \psi_3 \psi_5 \\
\theta_{4}  &=  \psi_5 - \psi_2 \psi_3  \psi_5 + \psi_1 \psi_3 \psi_4\\
\theta_{5} &=   \psi_1 \psi_6  - 2 \psi_1  \psi_2 \psi_3 \psi_6 - \psi_1^2 \psi_3 \psi_7 -  \psi_2 \psi_7 +  \psi_2^2 \psi_3 \psi_7 \\
\theta_{6} &=  \psi_1 \psi_7 - 2 \psi_1  \psi_2 \psi_3 \psi_7 + \psi_1^2 \psi_3 \psi_6 +   \psi_2 \psi_6  - \psi_2^2 \psi_3 \psi_6\\
\theta_{7}  &= \psi_1\psi_8  - \psi_2 \psi_9 \\
\theta_{8}  &= \psi_1\psi_9 + \psi_2 \psi_8.
\end{aligned}
\label{theta2psi}
\end{equation}

Eq. \eqref{theta2psi} can be written in a matrix form for branch $p$-$q$ as: 
\begin{equation}
\label{Theta_psi_relation_abstraction}
\begin{aligned}
\begin{bmatrix}
\theta_{pq}
\end{bmatrix} = 
\begin{bmatrix}
f(\psi_{pq})
\end{bmatrix},
\end{aligned}
\end{equation}
where $f$ is a nonlinear function.
Substitute \eqref{Theta_psi_relation_abstraction} in \eqref{Abstracted_SLIC_linear} to get:
\begin{equation}
\label{D_SLIC_abstraction}
\begin{aligned}
\begin{bmatrix}
D_{pq} +  \Delta D_{pq}
\end{bmatrix}
\begin{bmatrix}
f(\psi_{pq})
\end{bmatrix}
=
\begin{bmatrix}
c_{pq} +  \Delta c_{pq}
\end{bmatrix}.
\end{aligned}
\end{equation}

In contrast to the system of linear equations in
\eqref{Abstracted_SLIC_linear}, the system of nonlinear equations in \eqref{D_SLIC_abstraction} model the unknown parameters \textit{directly} within the problem set-up. Henceforth, \eqref{D_SLIC_abstraction}
is 
referred to as the nonlinear-SLIC (NL-SLIC) 
formulation.

\vspace{-1em}

\subsection{Optimization Formulation based on RQMs' Properties}
\label{Incorporating_domain_knowledge_SLIC}

The unknown vector $\psi_{pq}$ in \eqref{psi_Definition_9dim} is estimated from \eqref{D_SLIC_abstraction} by solving the following minimization problem 
that treats the additive noise as an unknown but small quantity:
\begin{equation}
\label{D_SLIC_initial_minimization}
\begin{aligned}
\arg\min_{\psi_{pq}} 
\left\| \begin{bmatrix}
D_{pq} 
\end{bmatrix} \begin{bmatrix}
f(\psi_{pq})
\end{bmatrix} - \begin{bmatrix}
 c_{pq} 
\end{bmatrix} \right\|_2^2 .
\end{aligned}
\end{equation}

However, due to the mismatch between the number of unknowns and equations, the problem is still ill-posed implying that a gradient descent-based solution of \eqref{D_SLIC_initial_minimization} will result in a non-unique answer.
To overcome this issue, we make use of the fact that the $6^{th}$ and $7^{th}$ parameters of \eqref{psi_Definition_9dim} are the real and imaginary parts of the ratio of correction factors of the CT and VT placed at the $p$-end of branch $p$-$q$.
Now, as the RQM-VT and RQM-CT were located at the $p$-end (see Section \ref{SLIC_modeling_section}) and RQMs have much smaller ratio errors than regular ITs, 
it can be surmised that $\psi_{pq}(6) \approx 1$ and $\psi_{pq}(7) \approx 0$.
Adding this as a regularizer term to \eqref{D_SLIC_initial_minimization} results in the optimization problem:
\begin{equation}
\label{D_SLIC_RQM_branch_with_Regularizer}
\begin{aligned}
\arg\min_{\psi_{pq}} \:\: & \left\| \begin{bmatrix} D_{pq} \end{bmatrix}    \begin{bmatrix} f(\psi_{pq}) \end{bmatrix} - \begin{bmatrix} c_{pq}   \end{bmatrix}  \right\|_2^2 + \lambda \left\|L \psi_{pq} - R \right\|_2^2,
\end{aligned}
\end{equation}
where
$\lambda$ is a tunable parameter,
{\color{black} and the matrix $L$ and vector $R$}, respectively, are 
given by:
\begin{equation}
    L [ i, j]= 
\begin{cases}
    1,& \text{if } i=j=6 \: \text{and} \: i=j=7\\
    0,              & \text{otherwise}.
\end{cases}
\label{L_dagger}
\end{equation}
\begin{equation}
    R [i]= 
\begin{cases}
    1,& \text{if } i=6\\
    0,              & \text{otherwise}.
\end{cases}
\label{R_dagger}
\end{equation}

Note that \eqref{D_SLIC_RQM_branch_with_Regularizer}-\eqref{R_dagger} holds only
when the SLIC problem for the RQM branch is expressed in terms of the physical parameters (denoted by $\psi_{pq}$), highlighting the advantage of the NL-SLIC formulation derived in \eqref{D_SLIC_abstraction} over the linear form of \eqref{Abstracted_SLIC_linear}.
At the same time, it is also true that
\eqref{D_SLIC_RQM_branch_with_Regularizer}-\eqref{R_dagger} is only applicable to the RQM branch, and since RQMs are expensive, it is unlikely that every branch will have 
an RQM pair at one end.
In the next section,
we build on
the formulation shown in \eqref{D_SLIC_RQM_branch_with_Regularizer}-\eqref{R_dagger} to solve
the NL-SLIC problem for non-RQM branches by exploiting power system domain knowledge. 

\section{Constrained Optimization Formulation for Solving SLIC Problem for Non-RQM Branches}
\label{SLIC-Non-RQM}

Before discussing the solution strategy for performing SLIC for non-RQM branches, 
we provide two insights in the form of
remarks.

\begin{remark}
So far, the IT CFRs of RQM branch $p$-$q$ were referenced to the VT at $p$-end of the line.
However, the SLIC parameter vector can be defined by changing the reference of the CFRs to $q$-end of branch $p$-$q$, 
which will cause $\alpha_{qp}$ to be in the denominator of the CFRs.\footnote{Note that this is a mathematical transformation only.
The location of the actual RQMs do not change.} 
In this case, we define $\psi_{qp}$ as the SLIC parameter vector, and the transformation function required to convert $\psi_{pq}$ to $\psi_{qp}$ will be
as follows:
\begin{equation}
\label{SLIC_parameter_Reference_swapping_within_branch}
\begin{aligned}
\begin{bmatrix}\psi_{qp} (1) \\ \psi_{qp} (2) \\ \psi_{qp} (3) \\ \psi_{qp} (4) \\ \psi_{qp} (5) \\ \psi_{qp} (6) \\ \psi_{qp} (7) \\ \psi_{qp} (8) \\ \psi_{qp} (9) \\ \end{bmatrix} = \begin{bmatrix}\psi_{pq} (1) \\ \psi_{pq} (2) \\ \psi_{pq} (3) \\ \mathcal{R}e\left[\frac{1}{\psi_{pq} (4) + j \psi_{pq} (5)}\right] \\ \mathcal{I}m\left[\frac{1}{\psi_{pq} (4) + j \psi_{pq} (5)}\right] \\ \mathcal{R}e\left[\frac{\psi_{pq} (6) + j \psi_{pq} (7)}{\psi_{pq} (4) + j \psi_{pq} (5)}\right] \\ \mathcal{I}m\left[\frac{\psi_{pq} (6) + j \psi_{pq} (7)}{\psi_{pq} (4) + j \psi_{pq} (5)}\right] \\\mathcal{R}e\left[\frac{\psi_{pq} (8) + j \psi_{pq} (9)}{\psi_{pq} (4) + j \psi_{pq} (5)}\right] \\ \mathcal{I}m\left[\frac{\psi_{pq} (8) + j \psi_{pq} (9)}{\psi_{pq} (4) + j \psi_{pq} (5)}\right] \\ \end{bmatrix}.
\end{aligned}
\end{equation} 
\end{remark}
\begin{remark}
The natural tendency of power utilities is to place PMUs on the highest voltage (HV) buses first \cite{varghese2024deep}. Therefore, the HV network becomes the prime candidate for the SLIC problem since it requires PMUs monitoring both ends of the lines. 
As such,
we define the term ``connected tree" as 
a network of HV lines (branches) such
that every bus is accessed from every other bus of the tree via the 
branches of the tree, 
and each line is monitored by PMUs placed on both of its ends.
Then, by definition, the RQM branch will be one branch that is present inside a connected tree. 
\end{remark}

\begin{figure}[t]
    \centering
    \includegraphics[width=\linewidth]{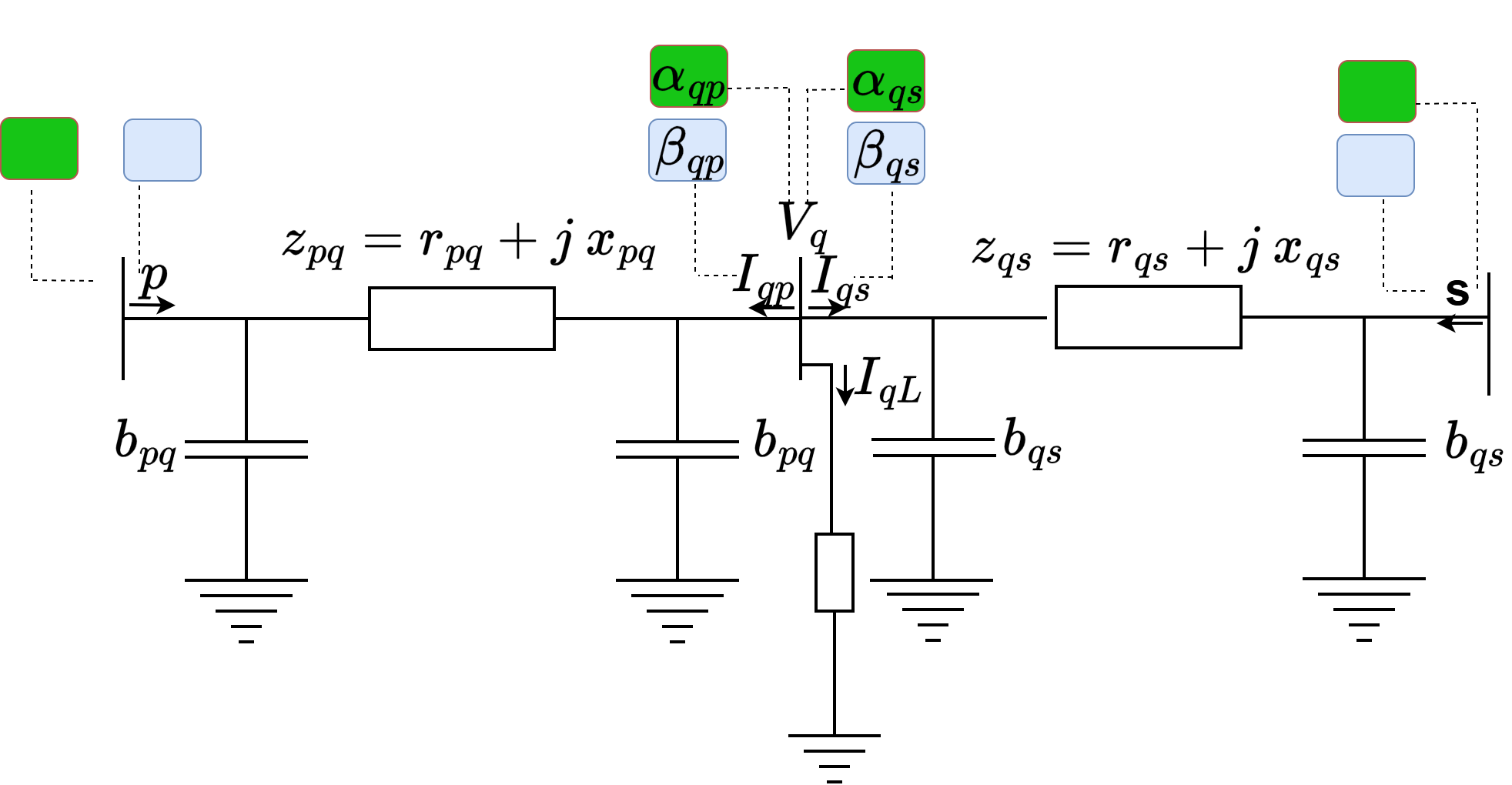} 
    \caption{$\pi$-model of two transmission lines connected to bus $q$}
    \label{KCL_eg_two_branches}
    \vspace{-1em}
\end{figure}

These two insights, related to transformation of $\psi$ and presence of RQM branch in a connected tree,
are crucial for creating
a new formulation that employs power system domain knowledge to solve
the NL-SLIC problem for non-RQM branches. 
We start 
by considering a branch $q$-$s$ that lies next to the RQM branch $p$-$q$ as shown in Fig. \ref{KCL_eg_two_branches}.

\vspace{-0.5em}
\subsection{Relation Connecting VTs Across a Bus}
\label{VoltRe}

Since the bus $q$ 
between branches $p$-$q$ and $q$-$s$
is common to 
both the branches, 
the true value of voltage at bus $q$ (i.e., $V_q^*$) is the same irrespective of which VT is used to measure it.
This redundancy in voltage measurement  
is combined
with the composite noise model (see Section \ref{SLIC_modeling_section}) to express $V_q^*$ in two ways as shown below:
\begin{equation}
\label{eqn:Vq-redundancy}
         \begin{aligned}
        V_q^* &= \alpha_{qp} (V_{qp} - \Delta V_{qp})  \\
        V_q^* &= \alpha_{qs} (V_{qs} - \Delta V_{qs}).  \\
         \end{aligned}
\end{equation}

Now, we define the variable $\rho_{pqs}$: 
\begin{equation}
\label{eqn:rho_definition}
         \begin{aligned}
        \rho_{pqs} = \frac{(V_{qp} - \Delta V_{qp})}{(V_{qs} - \Delta V_{qs})} = \frac{\alpha_{qs}}{\alpha_{qp}}. \\
         \end{aligned}
\end{equation}
where
$ \rho_{pqs}$ is the ratio of the correction factors of the VTs located at the $q$-end of branches $p$-$q$ and $q$-$s$.
An estimate of $\rho_{pqs}$, denoted by $\hat{\rho}_{pqs}$, can be obtained from $N$ noisy \textit{historical} PMU measurements ($V^{\mathrm{Hist}}_{qp}$, $V^{\mathrm{Hist}}_{qs}$) as shown below:
\begin{equation}
\label{eqn:rho_realistic_from_noisy_meas}
         \begin{aligned}
       \hat{\rho}_{pqs} = \left( \frac{ \sum\limits_{j=1}^N  V^{\mathrm{Hist}}_{qp}(j)}{\sum\limits_{j=1}^N  V^{\mathrm{Hist}}_{qs}(j)} \right).
         \end{aligned}
\end{equation}

In \eqref{eqn:rho_realistic_from_noisy_meas}, we leverage the fact that (a) calibration is always performed using ambient (non-fault) data, and (b) noise added by the PMU device is small and has a zero-mean Gaussian distribution \cite{wang2019transmission}. Thus, \eqref{eqn:rho_realistic_from_noisy_meas}  
captures the relation between the VT correction factors of two neighboring branches.

\vspace{-1em}
\subsection{Relation Connecting CTs Across a Bus}

Applying Kirchhoff's current law (KCL) at bus $q$ of Fig. \ref{KCL_eg_two_branches} yields the following equation when expressed using non-erroneous measurements:
\begin{equation}
\label{Ideal_KCL_at_Q}
    \begin{aligned}
 I_{qp}^* +  I_{qs}^* +  I_{qL}^* &= 0,
    \end{aligned}
\end{equation}
where $I_{qp}^*$ and  $I_{qs}^*$ denote the complex currents flowing into $q$ via branches $p$-$q$ and $q$-$s$, respectively. The sum of all other currents coming into $q$ is denoted using $I_{qL}^*$.
By applying the composite noise model, 
\eqref{Ideal_KCL_at_Q} can be re-written as:
\begin{equation}
\label{Realistic_KCL_at_Q}
    \begin{aligned}
 \beta_{qp} (I_{qp} - \Delta I_{qp} ) + \beta_{qs} (I_{qs}  - \Delta I_{qp}) + \beta_{qL} (I_{qL}  - \Delta I_{qp}) &= 0.
    \end{aligned}
\end{equation}

{\color{black} Dividing \eqref{Realistic_KCL_at_Q} by $\beta_{qp} \ne 0$, and performing
simple algebraic manipulations, we get: 
\begin{equation}
\label{EIVform}
    \begin{aligned}
\gamma_{pqs} (I_{qs}  - \Delta I_{qp}) + \gamma_{pqL} (I_{qL}  - \Delta I_{qp}) &=  -(I_{qp} - \Delta I_{qp} ),
    \end{aligned}
\end{equation}
where $ \gamma_{pqs} = \frac{\beta_{qs}}{ \beta_{qp}}$ and $ \gamma_{pqL} = \frac{\beta_{qL}}{ \beta_{qp}}$. Eq. \eqref{EIVform} can be expressed as a linear model 
in the errors-in-variables form as:
\begin{align}\label{EIVform2}
     \begin{bmatrix} I_{qs} -\Delta I_{qs}  & I_{qL}  - \Delta I_{qL} \end{bmatrix} \begin{bmatrix}  \gamma_{pqs} \\   \gamma_{pqL} \end{bmatrix} =  - \begin{bmatrix} I_{qp}  - \Delta I_{qp}   \end{bmatrix}. 
\end{align}

We can now estimate the ratio of the correction factors of the CTs by solving the following optimization problem:
\begin{align}
 \hspace{-3.0mm}   \begin{split}
\label{KCL_at_Q_n_time_instants_complex_T1}
\underset{\gamma_{pqs},\, \gamma_{pqL},\, \Delta I_{qp},\, \Delta I_{qs},\, \Delta I_{qL}}{\text{arg min}}
\quad &
\left\|
    \begin{bmatrix}
        \Delta I_{qs} & \Delta I_{qL}
    \end{bmatrix}
    \;
    \begin{bmatrix}
        \Delta I_{qp}
    \end{bmatrix}
\right\|_{F} \\
\text{subject to} \quad & \eqref{EIVform2}       
    \end{split}
\end{align}}
where $\|\|_F$ is the Frobenius norm. 
Our goal is to estimate
$\gamma_{pqs}$, which is the ratio of the correction factors of CTs located at $q$-end of branches $p$-$q$ and $q$-$s$.
Since the PMU noise 
is small and Gaussian \cite{wang2019transmission}, an
estimate of $\gamma_{pqs}$, denoted by $\hat{\gamma}_{pqs}$, can be obtained 
by solving \eqref{KCL_at_Q_n_time_instants_complex_T1} as a total least squares (TLS) problem \cite{markovsky2007overview} using \textit{historical} PMU measurements.

\vspace{-1em}
\subsection{Equality Constrained Formulation for Solving the NL-SLIC Problem for Consecutive Branch Pairs}

Using 
the estimates of $\rho_{pqs}$ and $\gamma_{pqs}$, 
the SLIC parameters of the branches $p$-$q$ and $q$-$s$ can be 
estimated simultaneously.
To that end, 
we 
establish a relationship between $\rho_{pqs}$ and $\gamma_{pqs}$: 
\begin{equation}
\label{rho_based_KCL_set_A_two_branch_example}
\begin{aligned}
 & \rho_{pqs} =\frac{\alpha_{qs}}{\alpha_{qp}} = \frac{\frac{\beta_{qs}}{\alpha_{qp}} }{ \frac{\beta_{qs}}{\alpha_{qs}} } = \frac{ \frac{\beta_{qs}}{\beta_{qp}} \frac{\beta_{qp}}{\alpha_{qp}} }{ \frac{\beta_{qs}}{\alpha_{qs}} } =  \frac{ \gamma_{pqs} \frac{\beta_{qp}}{\alpha_{qp}} }{ \frac{\beta_{qs}}{\alpha_{qs}} } \\
 &\iff   \rho_{pqs}  \frac{\beta_{qs}}{\alpha_{qs}} = \gamma_{pqs} \frac{\beta_{qp}}{\alpha_{qp}} \\
 &\iff   \gamma_{pqs} (\psi_{{qp}} (6) + j \psi_{{qp}} (7)) - \rho_{pqs}  (\psi_{{qs}} (6) + j \psi_{{qs}} (7))  = 0.
\end{aligned}
\end{equation}

In \eqref{rho_based_KCL_set_A_two_branch_example}, $\psi_{qp}$ denotes the SLIC parameters for branch $p$-$q$ referenced to the VT at the $q$-end, whereas $\psi_{qs}$ denotes the SLIC parameters for the branch $q$-$s$.
By using the estimates $\hat{\rho}_{pqs}$ and $\hat{\gamma}_{pqs}$ from \eqref{eqn:rho_realistic_from_noisy_meas} and \eqref{KCL_at_Q_n_time_instants_complex_T1}, respectively, and by separating real and imaginary parts of \eqref{rho_based_KCL_set_A_two_branch_example}, the following 
equality constraint is derived:
\begin{equation}
\begin{aligned}
\label{KCL_Gamma_Rho_Linear_EC}
&A_{pqs}  
   \begin{bmatrix} \psi_{{qp}} (6) \\ \psi_{{qp}} (7) \\\psi_{{qs}} (6) \\ \psi_{{qs}} (7) \end{bmatrix} = 0,
   \text{where }  \\
   & A_{pqs}\!=\begin{bmatrix}  \mathcal{R}e(\hat{\gamma}_{pqs}) & -  \mathcal{I}m(\hat{\gamma}_{pqs}) &  - \mathcal{R}e(\hat{\rho}_{{pqs}}) &   \mathcal{I}m(\hat{\rho}_{{pqs}}) \\
 \mathcal{I}m(\hat{\gamma}_{pqs}) &  \mathcal{R}e(\hat{\gamma}_{pqs}) &  -  \mathcal{I}m(\hat{\rho}_{{pqs}}) & -  \mathcal{R}e(\hat{\rho}_{{pqs}}) 
  \end{bmatrix}.
\end{aligned}
\end{equation}

Finally, for branch pairs $p$-$q$ and $q$-$s$, an equality constrained optimization formulation for solving the NL-SLIC problem is obtained as shown below:
\begin{equation}
\label{D_SLIC_form_Two_branch_Eq_constr_Optimization}
\begin{aligned}
\arg\min_{\psi_{qp}, \psi_{qs}} \:\: &\underbrace{\left\| \begin{bmatrix} D_{qp} & 0 \\ 0 & D_{qs} \end{bmatrix}    \begin{bmatrix} f(\psi_{qp}) \\ f(\psi_{qs}) \end{bmatrix} - \begin{bmatrix} c_{qp} \\ c_{qs} \end{bmatrix}  \right\|_2^2}_{\triangleq g(\psi_{qp}, \psi_{qs})} \\
\text{subject to  }  & A_{pqs}
  \begin{bmatrix} \psi_{{qp}} (6) \\ \psi_{{qp}} (7) \\\psi_{{qs}} (6) \\ \psi_{{qs}} (7) \end{bmatrix} = 0.\\
\end{aligned}
\end{equation}

It is observed from \eqref{D_SLIC_form_Two_branch_Eq_constr_Optimization} that for any pair of consecutive branches,
the NL-SLIC problem can be framed as an equality constrained optimization problem where the objective function is a combination of objective functions of the individual branches and the equality constraint connecting the SLIC parameters as described in 
\eqref{KCL_Gamma_Rho_Linear_EC}. 
Next, we show how
\eqref{D_SLIC_form_Two_branch_Eq_constr_Optimization}
can be applied to the entire connected tree.

\subsection{Network-wide Solution to NL-SLIC Problem}
To solve the NL-SLIC problem for a connected tree, we first solve it for the RQM branch (using \eqref{D_SLIC_RQM_branch_with_Regularizer}-\eqref{R_dagger}), and then proceed by taking two lines
at a time starting from the RQM-branch and its immediate neighboring branch.
This enables us to add the SLIC parameter estimates from the previous branch as a regularization term to \eqref{D_SLIC_form_Two_branch_Eq_constr_Optimization}, by employing \eqref{SLIC_parameter_Reference_swapping_within_branch} in Remark $1$.
The regularizer
enhances the numerical stability of the resulting optimization formulation.
The final \textit{constrained regularized optimization model} is as follows:
\begin{equation}
\label{Branch_Pair_Wise_Eq_constr_Optimization}
\begin{aligned}
\arg\min_{\psi_{qp}, \psi_{qs}} \:\: & g(\psi_{qp}, \psi_{qs}) + \lambda_{1} \left\| \psi_{qp} -   \hat{\psi}_{qp} \right\|_2^2\\
\text{subject to  }  & A_{pqs}
  \begin{bmatrix} \psi_{{qp}} (6) \\ \psi_{{qp}} (7) \\\psi_{{qs}} (6) \\ \psi_{{qs}} (7) \end{bmatrix} = 0,
\end{aligned}
\end{equation}
where $g(\psi_{qp}, \psi_{qs})$ is the loss function in \eqref{D_SLIC_form_Two_branch_Eq_constr_Optimization} and 
$\lambda_{1}\geq0 $ is a tuning parameter
that allocates appropriate weightage to the regularization term. \textcolor{black}{Henceforth, we refer to $\lambda$ of \eqref{D_SLIC_RQM_branch_with_Regularizer} and $\lambda_1$ of \eqref{Branch_Pair_Wise_Eq_constr_Optimization} as regularization coefficients.
}

\vspace{-2em}

\textcolor{black}{
\subsection{Strategy Employed to Solve the Proposed Formulation}
\label{D_SLIC_Solution_NewEquiTrust_section}
 For solving the regularized optimization formulation for the RQM branch (namely, \eqref{D_SLIC_RQM_branch_with_Regularizer}-\eqref{R_dagger}), Newton's method is sufficient. However, for solving the equality-constrained regularized optimization problem for the non-RQM branches (namely, \eqref{Branch_Pair_Wise_Eq_constr_Optimization}), an equality-constrained Newton's method enhanced with a trust region is employed. The trust-region method modifies the Newton’s method by introducing a “trust region” around the current iterate at each iteration \cite{yuan2015recent}. Instead of directly implementing the Newton update step, the method solves a sub-problem where it minimizes a model of the objective function within this trust region. The size of the trust region is adjusted iteratively: if the model within this region accurately predicts a reduction in the objective function, the region is expanded; if it fails, the region is reduced. This adaptive adjustment enables the trust-region based method to effectively navigate complex optimization landscapes.
As the proposed approach employs the \underline{N}ewton's method with \underline{E}quality constraint modified using a \underline{T}rust region for solving the SLIC problem, this solution method is henceforth referred to as NewEquiTrust-SLIC or NET-SLIC for short.}

\vspace{-1em}

\subsection{Reconstructing IT Correction Factors from IT CFRs}
\label{NL_SLIC_IT_Reconstruction_step}

The line parameters and IT CFRs of the branches of the connected tree are obtained
by solving \eqref{Branch_Pair_Wise_Eq_constr_Optimization}. 
However, the 
eventual goal is to estimate the correction factors from the CFRs.
This is done by following a similar procedure in which the estimation is done first for the RQM branch and then for the non-RQM branches.
For the RQM  branch $p$-$q$, we can solve \eqref{D_SLIC_RQM_branch_with_Regularizer}-\eqref{R_dagger}
a large number of times (say, $M$), and take the average to calculate $\hat{\alpha}_{qp}$, $\hat{\beta}_{pq}$, and $\hat{\beta}_{qp}$ as shown below:
\begin{equation}
\label{RQM_branch_CFR_to_CF}
         \begin{aligned}
    \hat{\alpha}_{qp}    &= \frac{1}{M}  \sum_{j=1}^M \widehat{{\left( \frac{\alpha_{qp}}{\alpha_{pq}} \right)}_j} , \:
  \hat{\beta}_{pq}    &= \frac{1}{M}  \sum_{j=1}^M {\widehat{\left( \frac{\beta_{pq}}{\alpha_{pq}} \right)}_j},\\
 \hat{\beta}_{qp}    &= \frac{1}{M}  \sum_{j=1}^M {\widehat{\left( \frac{\beta_{qp}}{\alpha_{pq}} \right)}_j} .\\   
         \end{aligned}
\end{equation}

\begin{figure}[ht]
            \centering
            \includegraphics[width=0.45\textwidth]{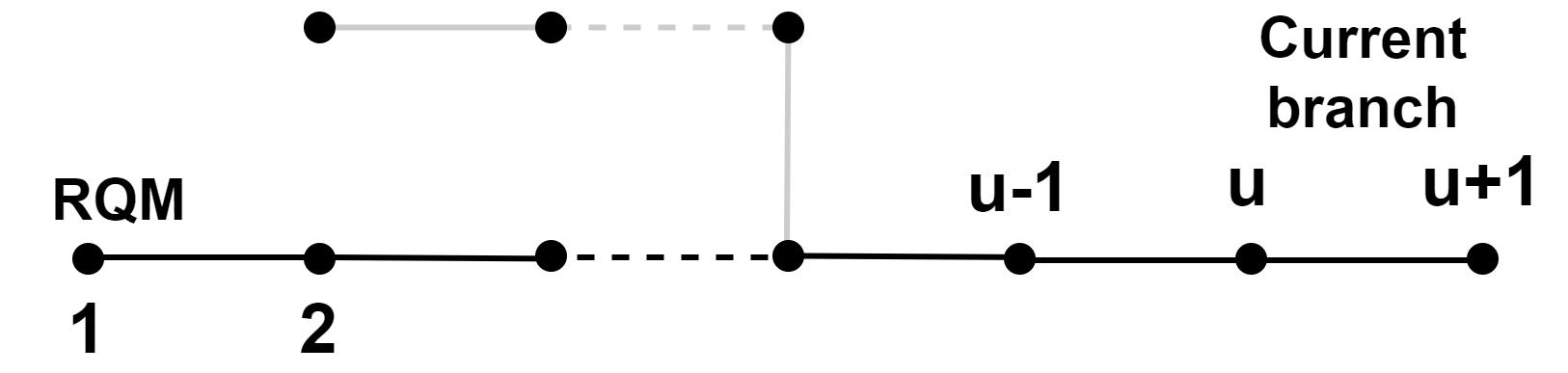}
            \caption{\textcolor{black}
            {Branches of the connected tree (in black) joining RQM branch to current branch; the lines in gray are not part of the connected tree}}
            \label{RQM_to_current_network_actiev_path}
            \vspace{-0.25em}
\end{figure}

Note that in \eqref{RQM_branch_CFR_to_CF} we used the fact that (a) IT correction factors change at a much slower rate than the speed with which PMUs produce measurements, and (b) RQMs have very small errors.  
For the non-RQM branches, first the path connecting the branch under consideration to the RQM branch in the connected tree is found as shown in Fig. \ref{RQM_to_current_network_actiev_path}.  
In the figure, $(1,2)$ is the RQM branch, and $(u,u+1)$ is the \textit{current branch} 
whose correction factors are to be estimated.
For $(u,u+1)$, the relation of CFRs to the RQM located at the $1$-end of branch $(1,2)$ is computed in the following way. First, calculate $\Lambda$ as:
\begin{equation}
\label{Lambda_for_generic_kappa}
\begin{aligned}
\Lambda &=\frac{1}{M} \sum_{j=1}^M \left( \hat{\rho}_{u_j} \frac{\hat{\alpha}_{(u,(u-1))_j}}{ \hat{\alpha}_{((u-1),u)_j}}   \hat{\rho}_{(u-1)_j} \dots  \frac{\hat{\alpha}_{(3,2)_j}}{\hat{\alpha}_{(2,3)_j}}   \hat{\rho}_{2_j} \frac{\hat{\alpha}_{(2,1)_j}}{\hat{\alpha}_{(1,2)_j}} \right)  \\
&= \frac{1}{M} \sum_{j=1}^M  \left( \left( \prod_{k=2}^{(u)} \hat{\rho}_{k_j} \right) \left( \prod_{(u, u+1)=(1,2)}^{((u-1), u))}  \frac{\hat{\alpha}_{(u+1,u)_j}}{\hat{\alpha}_{(u,u+1)_j}}  \right) \right).
\end{aligned}
\end{equation}

Then, the correction factors for $(u,u+1)$ is computed using $\Lambda$ and the CFRs of every line segment in between $(u,u+1)$
and the RQM branch $(1,2)$ as shown below:
\begin{equation}
\label{Generic_kappa_Estimation_HV_network}
\begin{aligned}
&\hat{\alpha}_{(u,u+1)} = \Lambda , \:\:
&\hat{\alpha}_{(u+1,u)} = \Lambda  \: \frac{\hat{\alpha}_{(u+1,u)}}{\hat{\alpha}_{(u,u+1)}} \\ 
&\hat{\beta}_{(u,u+1)} = \Lambda \: \frac{\hat{\beta}_{(u,u+1)}}{\hat{\alpha}_{(u,u+1)}}, \:\: 
&\hat{\beta}_{(u+1,u)} = \Lambda \: \frac{\hat{\beta}_{(u+1,u)}}{\hat{\alpha}_{(u,u+1)}}  . \\
\end{aligned}
\end{equation}

\textcolor{black}{Note that in \eqref{Generic_kappa_Estimation_HV_network}, the terms $\frac{\hat{\alpha}_{(u+1,u)}}{\hat{\alpha}_{(u,u+1)}}$, $\frac{\hat{\beta}_{(u,u+1)}}{\hat{\alpha}_{(u,u+1)}}$, and $\frac{\hat{\beta}_{(u+1,u)}}{\hat{\alpha}_{(u,u+1)}}$ denote the estimated CFRs, implying that each of them is a single complex number. When pre-multiplied with $\Lambda$, they provide estimates of the corresponding correction factors.}

The overall process for solving the NL-SLIC problem is summarized in Algorithm \ref{alg:Sequential_NET_SLIC}. Its inputs are the noisy voltage and current phasors of the connected tree  as well as the location of the RQM branch, while its outputs are the estimates of the line parameters and IT correction factors.
The estimation process begins by solving for the SLIC parameters of the RQM branch using Newton's method and \eqref{RQM_branch_CFR_to_CF} as seen in Steps 3-5.
For every non-RQM branch, 
the path denoted by $\zeta$, connecting that branch with the nearest RQM branch is determined.
Starting from the RQM branch, and taking one branch pair at a time, the SLIC parameters of each branch pair inside $\zeta$ are estimated using NET-SLIC; this is shown in Steps 10-13.  
Finally, the IT correction factors are 
recovered using Step 14.

\begin{algorithm}[ht]
\caption{Proposed Constrained Formulation for NL-SLIC} \label{alg:Sequential_NET_SLIC}
\begin{algorithmic}[1] 
\State \textbf{Input: }   

Connected Tree (${\mathcal{L}}$), $V_{\mathcal{L}}$, $I_{\mathcal{L}}$, $\mathrm{RQM}$ $\mathrm{Branch}$
\State \textbf{Output: }  $\hat{r}_{pq}, \hat{x}_{pq}, \hat{b}_{pq}, \hat{\alpha}_{pq}, \hat{\alpha}_{qp}, \hat{\beta}_{pq}, \hat{\beta}_{qp} \: \forall (p,q) \in \mathcal{L}$
    \ForAll 
    {$\mathrm{RQM}$ $\mathrm{Branch} \in \mathcal{L}$}
        \State 
        Solve \eqref{D_SLIC_RQM_branch_with_Regularizer}-\eqref{R_dagger} using Newton's method
        \State Recover IT correction factors using \eqref{RQM_branch_CFR_to_CF}
    \EndFor    
    \ForAll {$(p,q)   \in \left(\mathcal{L} \setminus \mathrm{RQM}\: \mathrm{Branch}
    \right)$}
        \State Find path $\zeta$ from $\mathrm{RQM}$ $\mathrm{Branch}$ to present branch 
        \State Set first branch in branch pair as previous branch in $\zeta$
        \State Set second branch in branch pair as present branch
        \State Calculate $\hat{\rho}_{pqs}$ using \eqref{eqn:rho_realistic_from_noisy_meas}
        \State Calculate $\hat{\gamma}_{pqs}$ using \eqref{KCL_at_Q_n_time_instants_complex_T1}
        \State Create equality constraint using \eqref{KCL_Gamma_Rho_Linear_EC}
        \State Create \eqref{Branch_Pair_Wise_Eq_constr_Optimization} and solve using NET-SLIC
        \State Recover IT correction factors 
        using \eqref{Lambda_for_generic_kappa} and \eqref{Generic_kappa_Estimation_HV_network}
    \EndFor
    \State \textbf{return} $\hat{r}_{pq}, \hat{x}_{pq}, \hat{b}_{pq}, \hat{\alpha}_{pq}, \hat{\alpha}_{qp}, \hat{\beta}_{pq}, \hat{\beta}_{qp}$
\end{algorithmic}
\end{algorithm}

\textcolor{black}{ In summary, the proposed approach leverages the presence of a pair of RQMs
within the connected tree network, along with Kirchhoff's current and voltage law-based equality constraints, to guide the solution towards a physically meaningful quantity.
The resulting optimization problem is \textit{nonlinear} with respect to the SLIC parameters. To tackle this problem, the constrained formulation augments the original objective function with a regularization term and a set of equality constraints. The regularization term serves as a soft prior that biases the solution toward a physically meaningful operating point - specifically, one in which the real parts of the correction factors of the RQMs
are close to unity and the imaginary parts are close to zero. Importantly, the regularization term does not rigidly fix these values to 1 and 0; rather, it gently penalizes deviations, thereby allowing the optimizer sufficient flexibility to converge to the true solution. Complementing this, the domain-knowledge-based equality constraints (derived from Kirchhoff's current and voltage laws) ensure that the final solution remains consistent with the underlying physics of the network.
In this way, the combination of the regularization term and the equality constraints collectively ensures that the obtained solution is both accurate and physically consistent. The Newton-based solver, operating under these modifications, converges to an acceptable
solution, as demonstrated in the next section.
}

\begin{remark}
\textcolor{black}{Note that the the focus of this paper is on positive sequence line parameter and correction factor estimation. With suitable modifications, (e.g., by representing each phase by its own set of line parameters and correction factors), the proposed formulation can be extended to three phase line parameter and IT correction factor estimation as well.}
\end{remark}

\section{Evaluating Performance of Proposed Approach for 
Performing SLIC
}
\label{Evaluating_NewEquiTrust_performance_section}

This section presents the results that were obtained when Algorithm \ref{alg:Sequential_NET_SLIC} was used to solve the NL-SLIC problem described in Sections \ref{SLIC-RQM} (for RQM branch) and \ref{SLIC-Non-RQM} (for non-RQM branches). 
The synthetic system used to evaluate performance was the IEEE 118-bus system (see Sections \ref{DataGen} and \ref{Simulated_Data_NETSLIC_results_IEEE118bus}). This system was also used to compare results with a state-of-the-art approach (in Section \ref{ComparisonSOTA}). The results obtained using 
field PMU data are described in Section \ref{Field}. The importance 
of performing SLIC for LSE is highlighted in Section \ref{LSE}.

\vspace{-0.5em}
\subsection{Synthetic Data Generation and Metrics for Evaluation}
\label{DataGen}

The true noise-free voltage and current phasor measurements for the 118-bus system were generated in MATPOWER by simulating the morning load pickup
in which the load increases by 60\% in one hour 
\cite{gao2012dynamic}.
\textcolor{black}{Next, $\eta$ for the regular ITs and RQMs were randomly selected from their ranges specified in  \cite{IEEE_C57_13_2016_std_for_ITs}, and multiplied with the true measurements. 
The regular ITs use the 0.6 accuracy class, while the RQMs are assigned an accuracy class of 0.15.
Finally, Gaussian noise having $0.1\%$ TVE \cite{pegoraro2022pmu} was added to produce the noisy measurement dataset.}

The proposed approach was implemented and evaluated using Python. For evaluating the accuracy of LPE, the absolute relative error ($\mathrm{ARE}$) metric was
used:
\begin{equation}
\label{ARE_basic_eq}
\mathrm{ARE (\%)} = \left|\frac{x_{\mathrm{est}} - x_{\mathrm{true}}}{x_{\mathrm{true}}}\right| \times 100,
\end{equation}
where $x_{\mathrm{est}}$ is the estimated parameter, and is $x_{\mathrm{true}}$ the true parameter. 
When performing large numbers of Monte Carlo (MC) simulations, the mean and standard deviation of $\mathrm{ARE}$ denoted by $\mathrm{MARE}$ and $\mathrm{SDARE}$, respectively, are calculated.
For the IT correction factors, which are complex numbers, the $\mathrm{ARE}$ metric is used for the magnitude component while the absolute error ($\mathrm{AE}$) metric is used for the angle component. The $\mathrm{AE}$ metric is mathematically expressed as:
\begin{equation}
\label{AE_basic_eq}
\mathrm{AE} = |x_{\mathrm{est}} - x_{\mathrm{true}}|.
\end{equation}

\begin{figure*}
            \centering
            \includegraphics[width=0.97\textwidth]{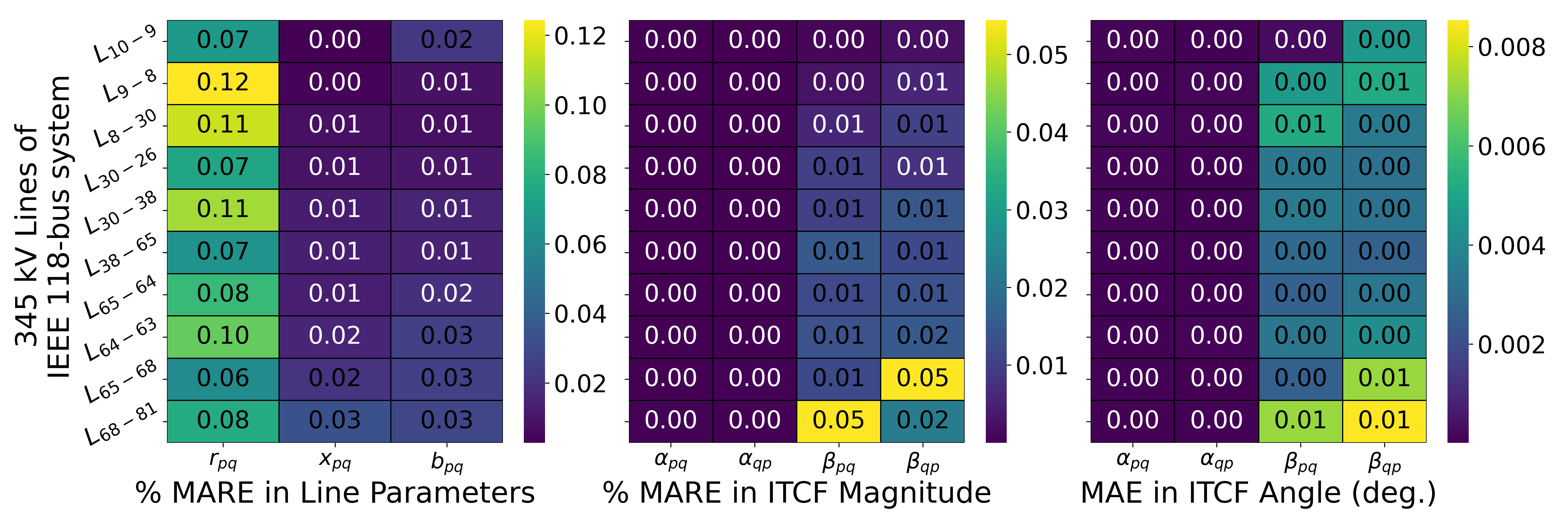} 
            \caption{Evaluation of line parameter and IT correction factor (CF) estimates in the ideal case (perfect RQM and no additive noise)}%
            \label{SLIC_IDEAL_Heatmap}
            \vspace{-0.5em}
\end{figure*}

\vspace{-1.5em}
\subsection{Results for IEEE 118-Bus System}
\label{Simulated_Data_NETSLIC_results_IEEE118bus}

 For evaluating the performance of the proposed approach, the 345 kV network
 of this system is utilized. It is assumed that PMUs are placed on all the 345 kV buses, and that they monitor every line that comes out of these buses (see Remark $2$ in Section \ref{SLIC-Non-RQM}). 
We first consider the ideal scenario in which $\eta$ for the RQMs is $1+j0$, and the additive noise is absent. \textcolor{black}{The regularization coefficients $\lambda$ and
$\lambda_1$ are set to $0.1$.}
The number of MC runs is $1000$ for every scenario, \textcolor{black}{and $n$ (see Section \ref{SLIC_modeling_section}) and $M$ (see Section \ref{NL_SLIC_IT_Reconstruction_step}) are assigned values of $60$ and $10$, respectively.} 

\looseness=-1 The results for the ideal scenario is displayed in Fig. \ref{SLIC_IDEAL_Heatmap}.
In this figure, the $\mathrm{MARE}$ for the line parameters are shown in the heatmap on the left, while 
the heatmaps depicting magnitude $\mathrm{MARE}$ and angle $\mathrm{MAE}$ for the IT correction factors are displayed 
in the center and right, respectively.
It is evident from the figure that all the SLIC parameters exhibit exceptionally low estimation errors.
The highest $\mathrm{MARE}$ for the line parameters is only $0.12\%$. 
For IT correction factors, the maximum magnitude $\mathrm{MARE}$ is $0.05\%$ and the maximum angle $\mathrm{MAE}$ is $0.008^{\circ}$.
These findings confirm that the proposed NET-SLIC gives extremely high accuracy under ideal conditions.

\begin{figure*}
    \centering
    \begin{subfigure}[b]{0.67\columnwidth}  
        \centering
        \includegraphics[width=\textwidth]{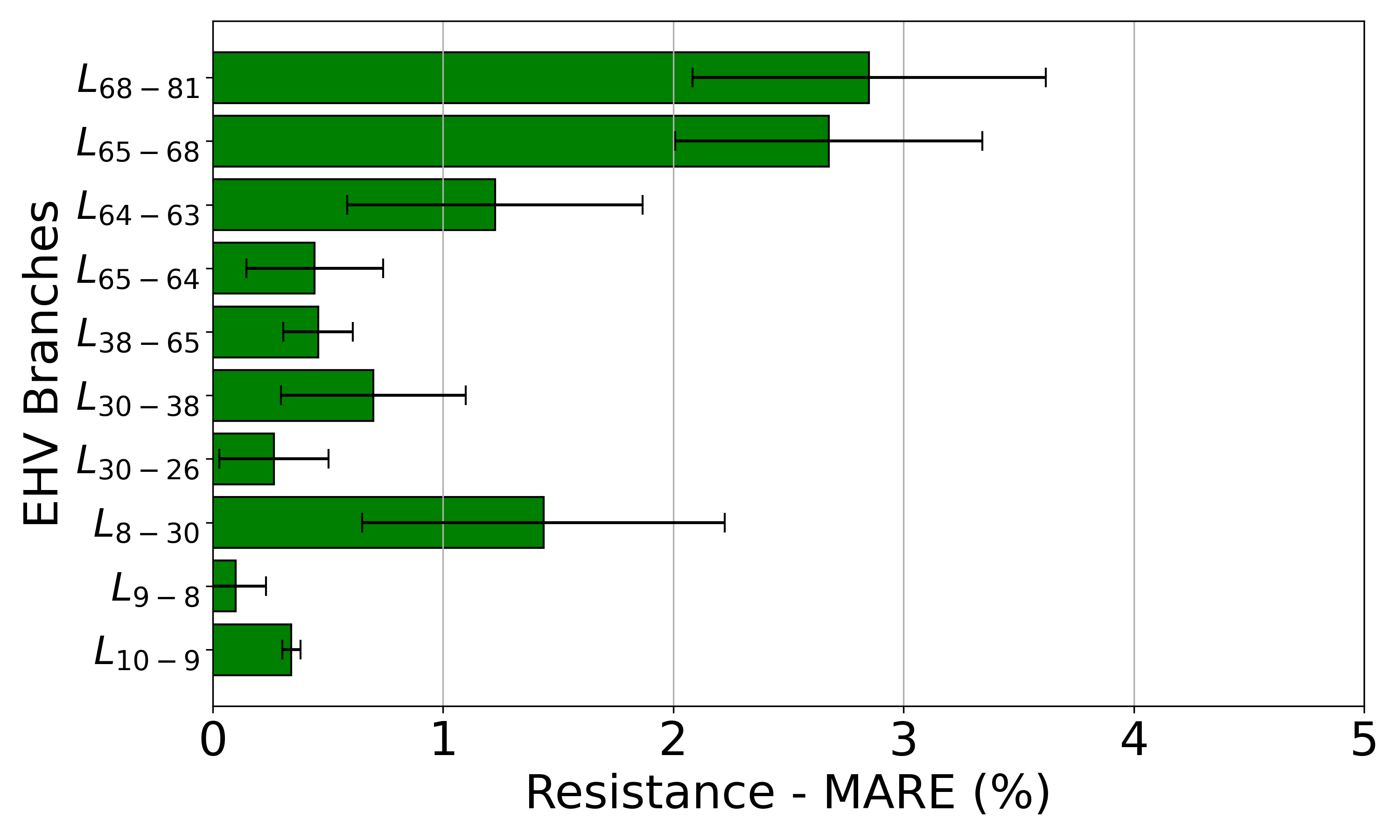}  
        \caption{Resistance }
        \label{C1_Resistance}
    \end{subfigure}
    \hfill
    \begin{subfigure}[b]{0.67\columnwidth}  
        \centering
        \includegraphics[width=\textwidth]{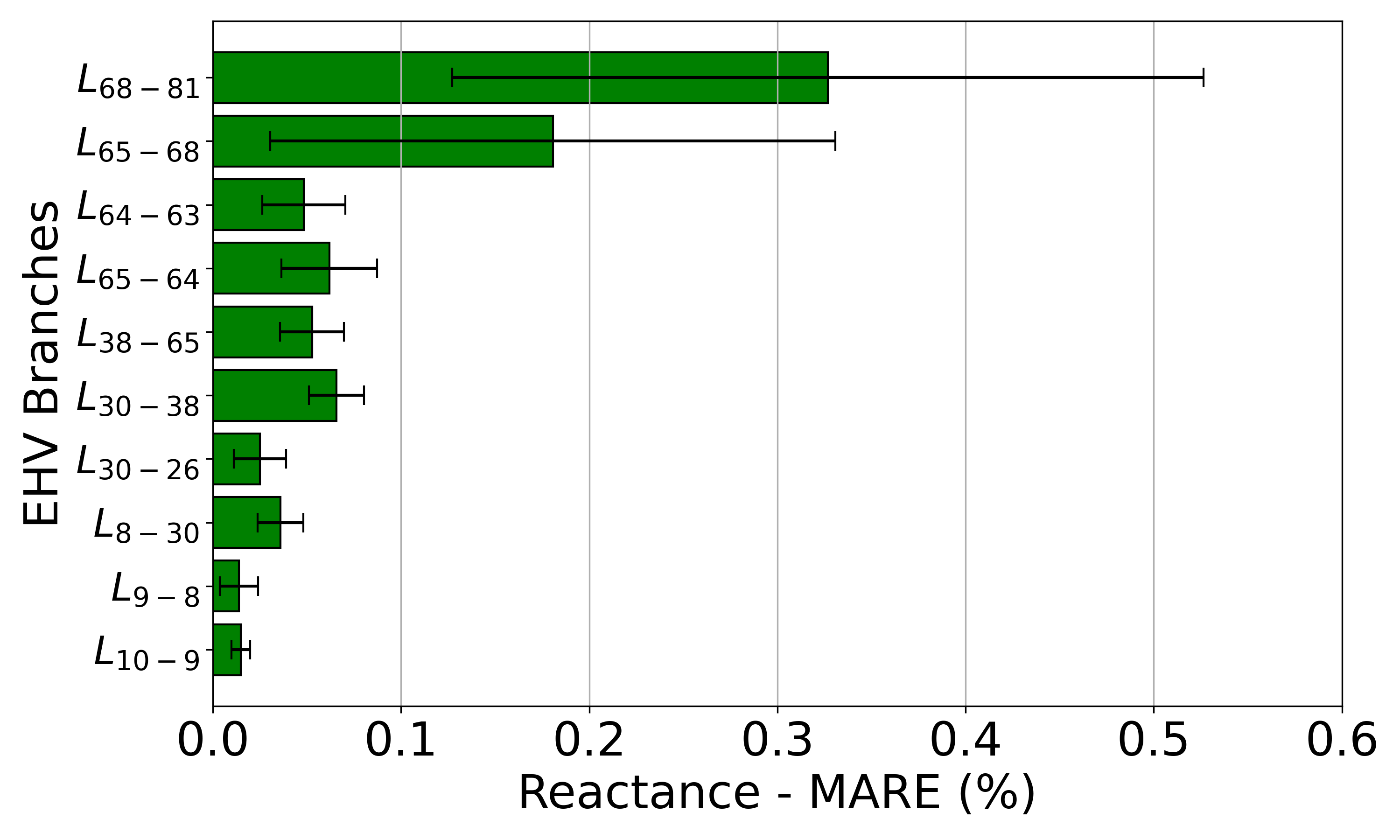} 
        \caption{Reactance }
        \label{C2_Reactance}
    \end{subfigure}
    \hfill
    \begin{subfigure}[b]{0.67\columnwidth}  
        \centering
        \includegraphics[width=\textwidth]{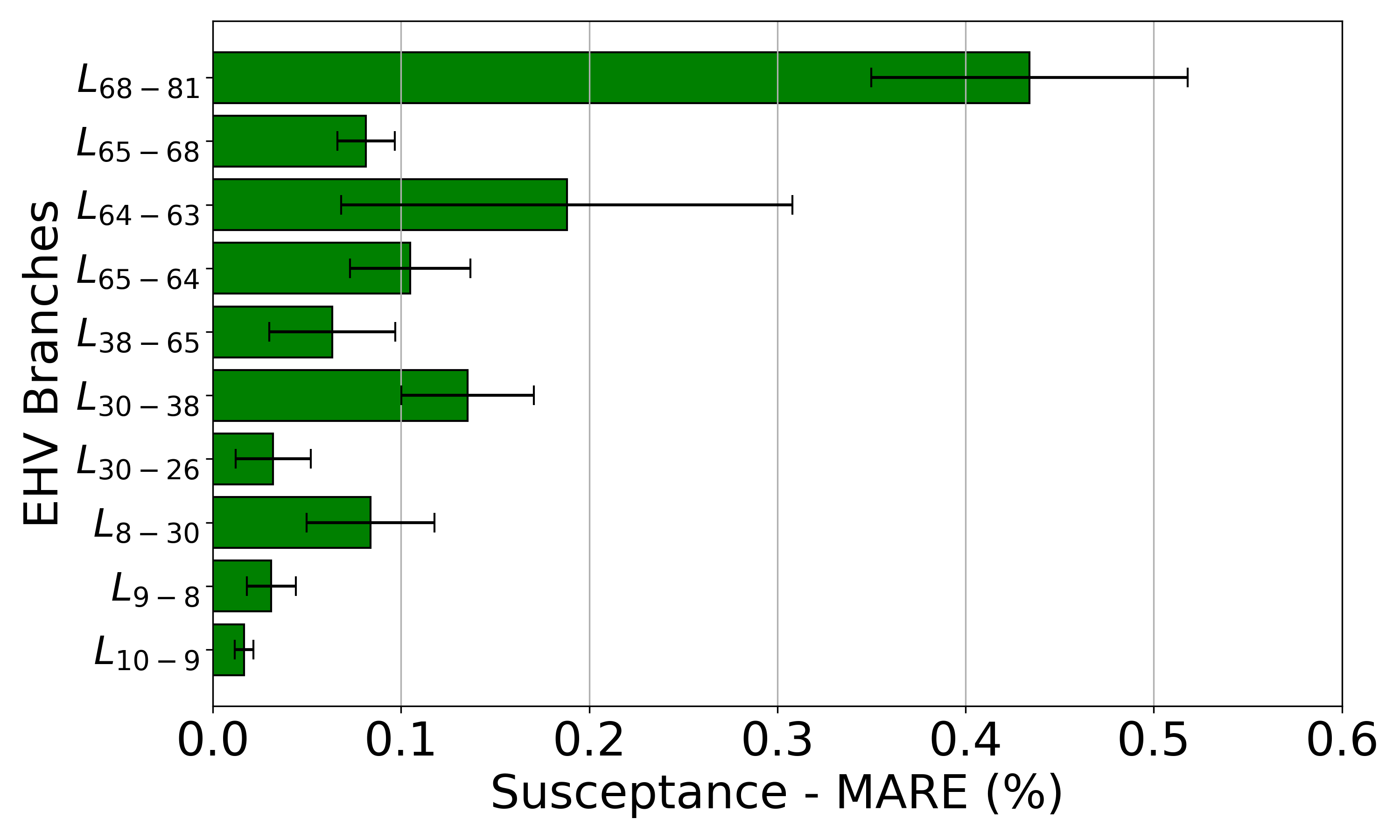} 
        \caption{Susceptance }
        \label{C2_Susceptance}
    \end{subfigure}
    \caption{Evaluation of line parameter estimates in presence of a perfect RQM pair but with additive PMU noise}
    \label{C2_SLIC_AddiNoise_Evaluation_of_Line_parameters}
            \vspace{-1em}
\end{figure*}

\begin{figure}[ht]
            \centering
            \vspace{-1em}
            \includegraphics[width=0.485\textwidth]{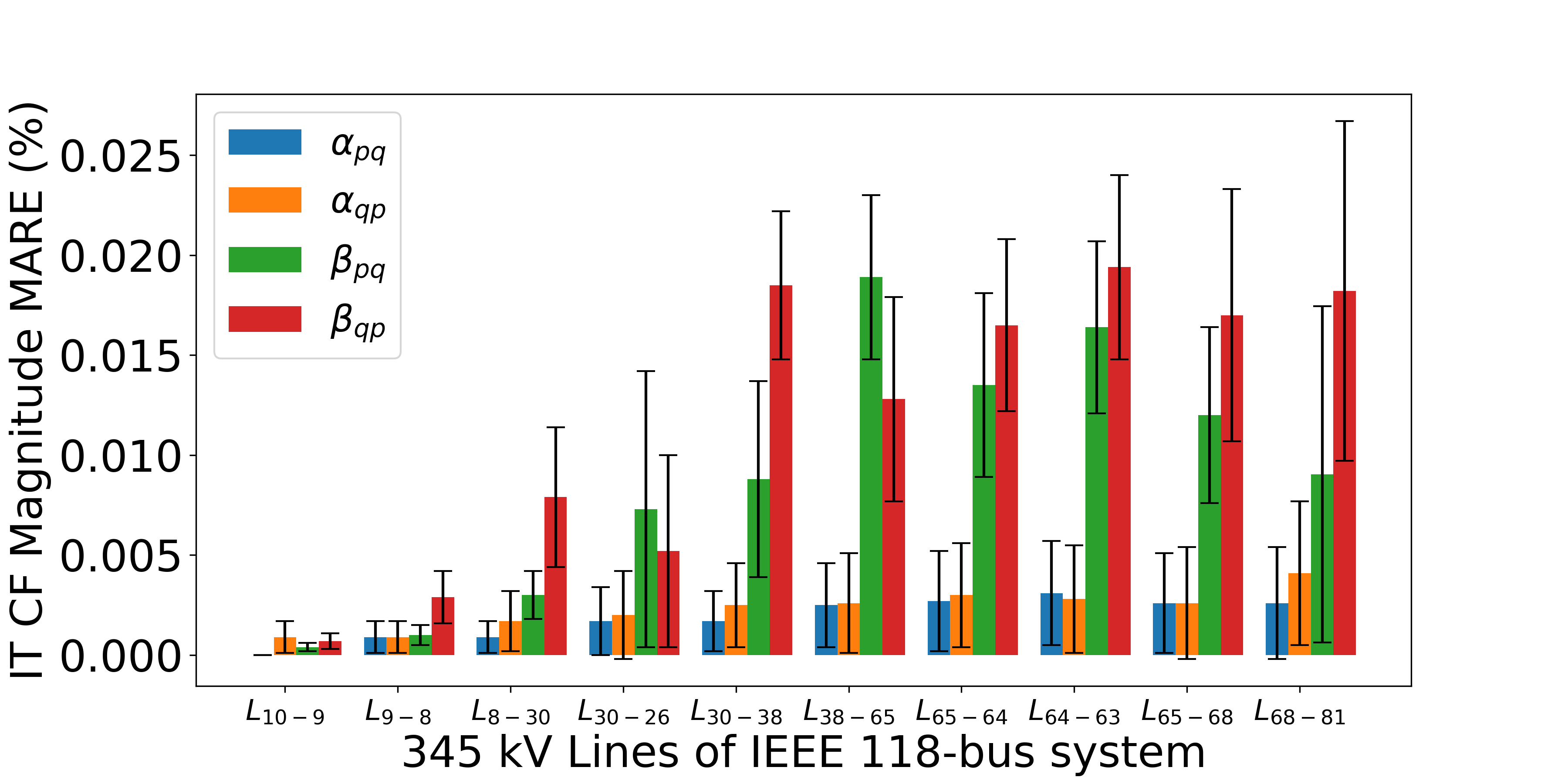}
            \caption{Accuracy of IT correction factor (CF) - magnitude, in presence of a perfect RQM pair but with additive PMU noise}%
            \label{C2_SLIC_AddiNoise_Evaluation_ITCF_Mag}
            \vspace{-1em}
\end{figure}

\begin{figure}[t]
            \centering
            \includegraphics[width=0.485\textwidth]{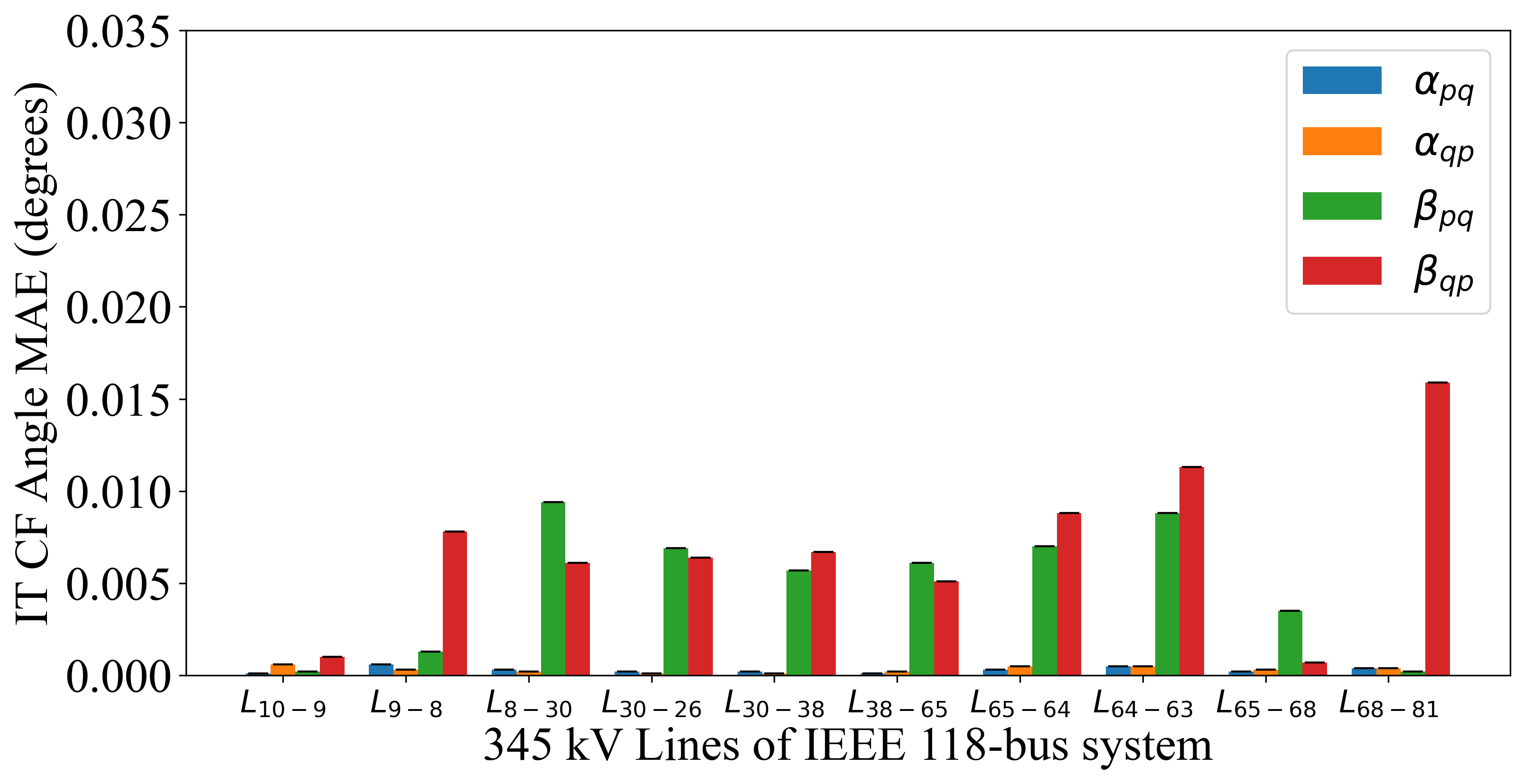}
            \caption{Accuracy of IT correction factor (CF) - angle, in presence of a perfect RQM pair but with additive PMU noise }%
            \label{C2_SLIC_AddiNoise_Evaluation_ITCF_Angle}
            \vspace{-1em}
\end{figure}

Next, the assumption of zero additive noise is relaxed by introducing additive noise with up to $0.1\%$ TVE in the PMU measurements. This scenario represents the 
case where the network has a well-calibrated IT set (a perfect RQM pair), 
but the measurements are affected by additive noise.
The $\mathrm{MARE}$ and $\mathrm{SDARE}$ of the line parameter estimates for this scenario is displayed in Fig. \ref{C2_SLIC_AddiNoise_Evaluation_of_Line_parameters}, while the accuracy of the IT correction factor estimates is provided in Figs.  \ref{C2_SLIC_AddiNoise_Evaluation_ITCF_Mag} and \ref{C2_SLIC_AddiNoise_Evaluation_ITCF_Angle}.

The results shown in Fig. \ref{C2_SLIC_AddiNoise_Evaluation_of_Line_parameters} indicate that the $\mathrm{MARE}$ for reactance and susceptance consistently remain below 1\%, indicating high estimation accuracy.
Even the variability of the $\mathrm{MARE}$, indicated by the $\mathrm{SDARE}$ and shown by the black whiskers on the bars, is also reasonable. 
The errors in the resistance parameter estimates is relatively high but much less than $5\%$ for all the lines.
One reason for this could be the comparatively small values of the resistances of transmission lines, which makes it harder for any estimation approach to give a low $\mathrm{ARE}$. 
These results confirm the robustness of the proposed approach for LPE in presence of PMU noise.

Fig. \ref{C2_SLIC_AddiNoise_Evaluation_ITCF_Mag} shows the estimation errors in the magnitudes of the IT correction factors for every branch. 
For a given branch, the first two bars represent the $\mathrm{MARE}$s for the VTs, while the last two correspond to the CTs. Similarly, the angle $\mathrm{MAE}$s for all four IT correction factor estimates are shown in Fig.  \ref{C2_SLIC_AddiNoise_Evaluation_ITCF_Angle}.
The IT correction factor estimates for the VTs is nearly perfect across all branches, with the magnitude $\mathrm{MARE}$ consistently below $0.002\%$ and the angle $\mathrm{MAE}$ less than $0.002^{\circ}$.
CT correction factor estimates are also accurate, with the \textit{maximum} $\mathrm{MARE}$ for CT correction factors being less than $0.020\%$ in magnitude and $0.02^{\circ}$ in angle, both of which are reasonably small quantities.
Therefore, if a pair of well-calibrated ITs is available at even a single location within the connected tree, the calibration of all the ITs in that connected tree can be successfully carried out in a highly accurate manner using the proposed approach.

\begin{figure*}
    \centering
    \begin{subfigure}[b]{0.67\columnwidth}  
        \centering
        \includegraphics[width=\textwidth]{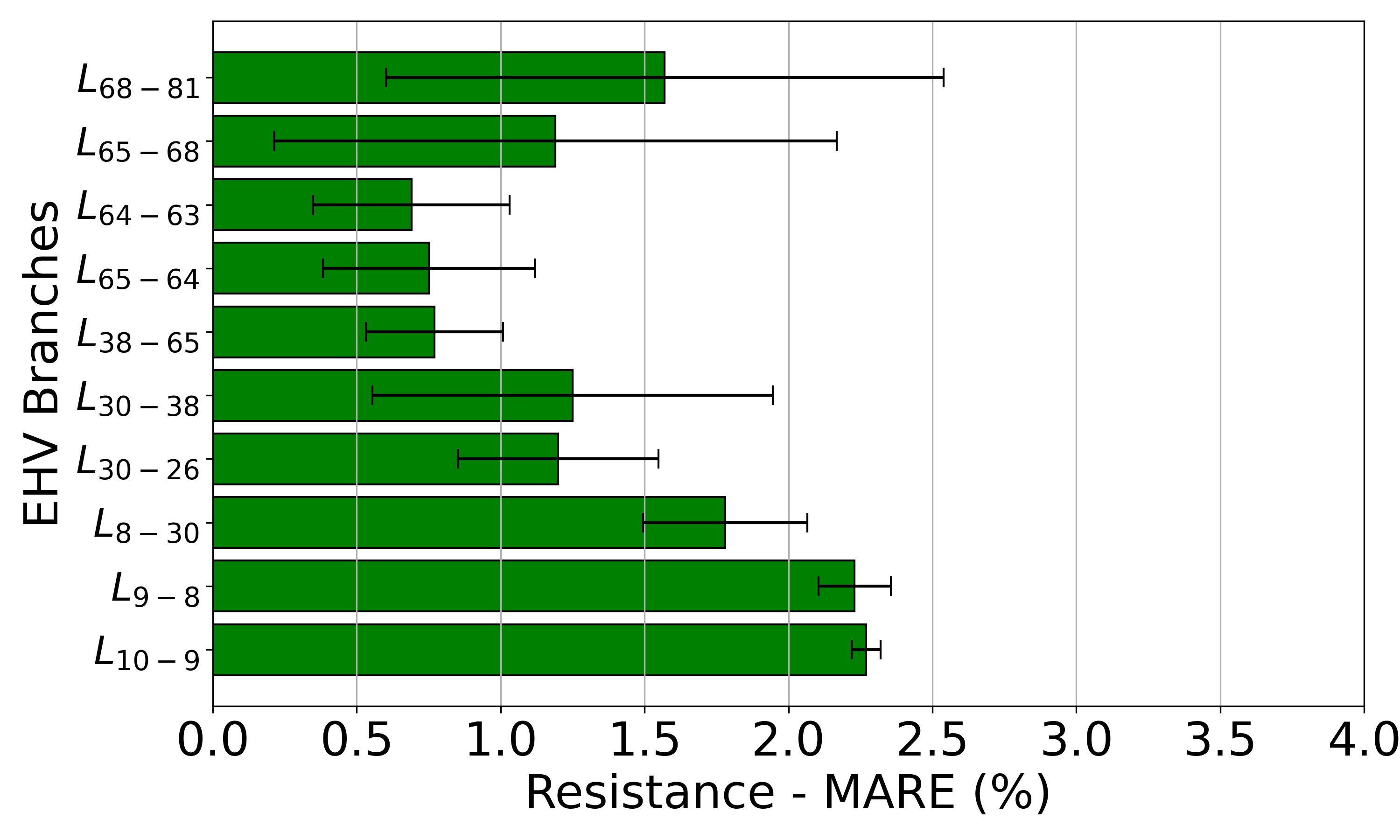}  
        \caption{Resistance }
        \label{C3_resistance}
    \end{subfigure}
    \hfill
    \begin{subfigure}[b]{0.67\columnwidth}  
        \centering
        \includegraphics[width=\textwidth]{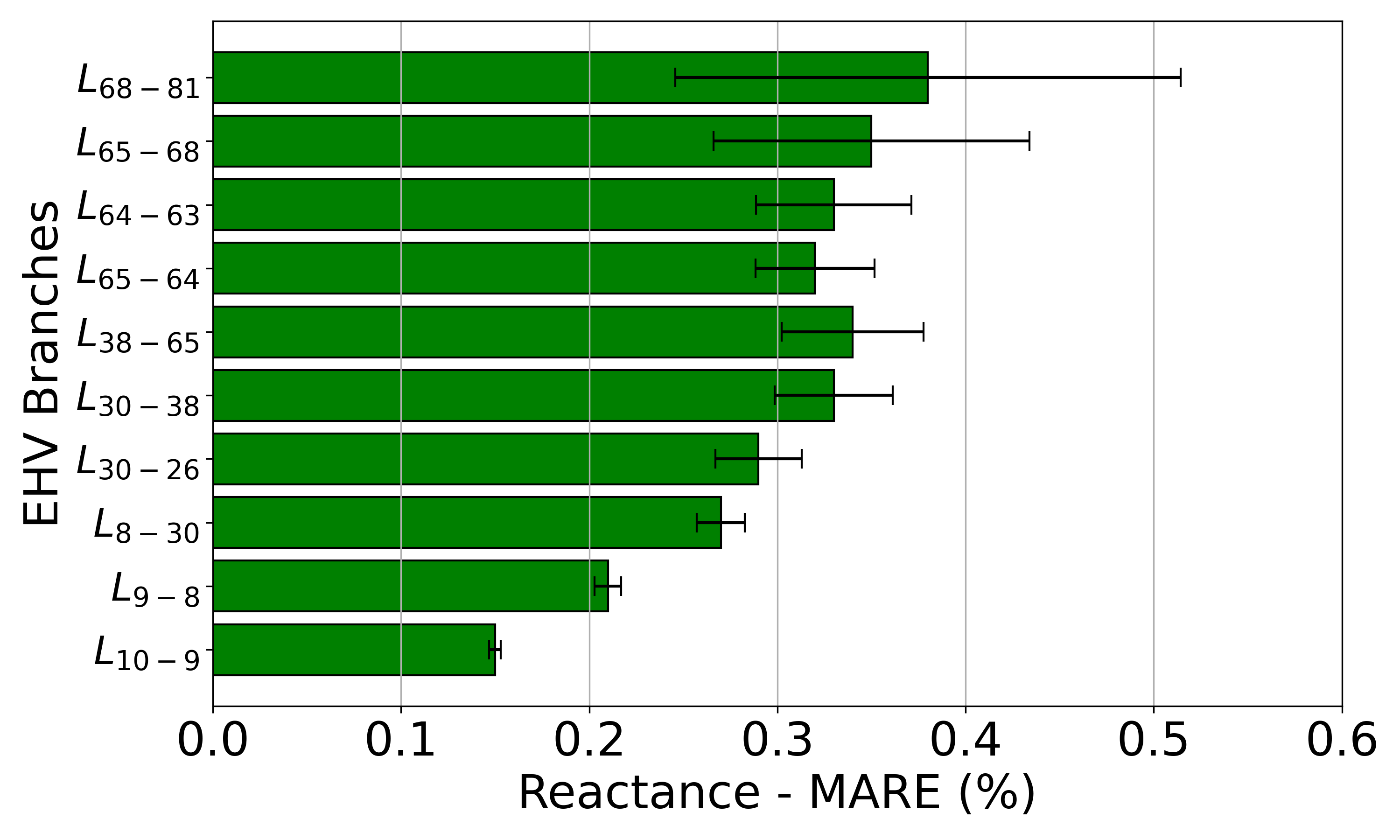} 
        \caption{Reactance }
        \label{C3_Reactance}
    \end{subfigure}
    \hfill
    \begin{subfigure}[b]{0.67\columnwidth}  
        \centering
        \includegraphics[width=\textwidth]{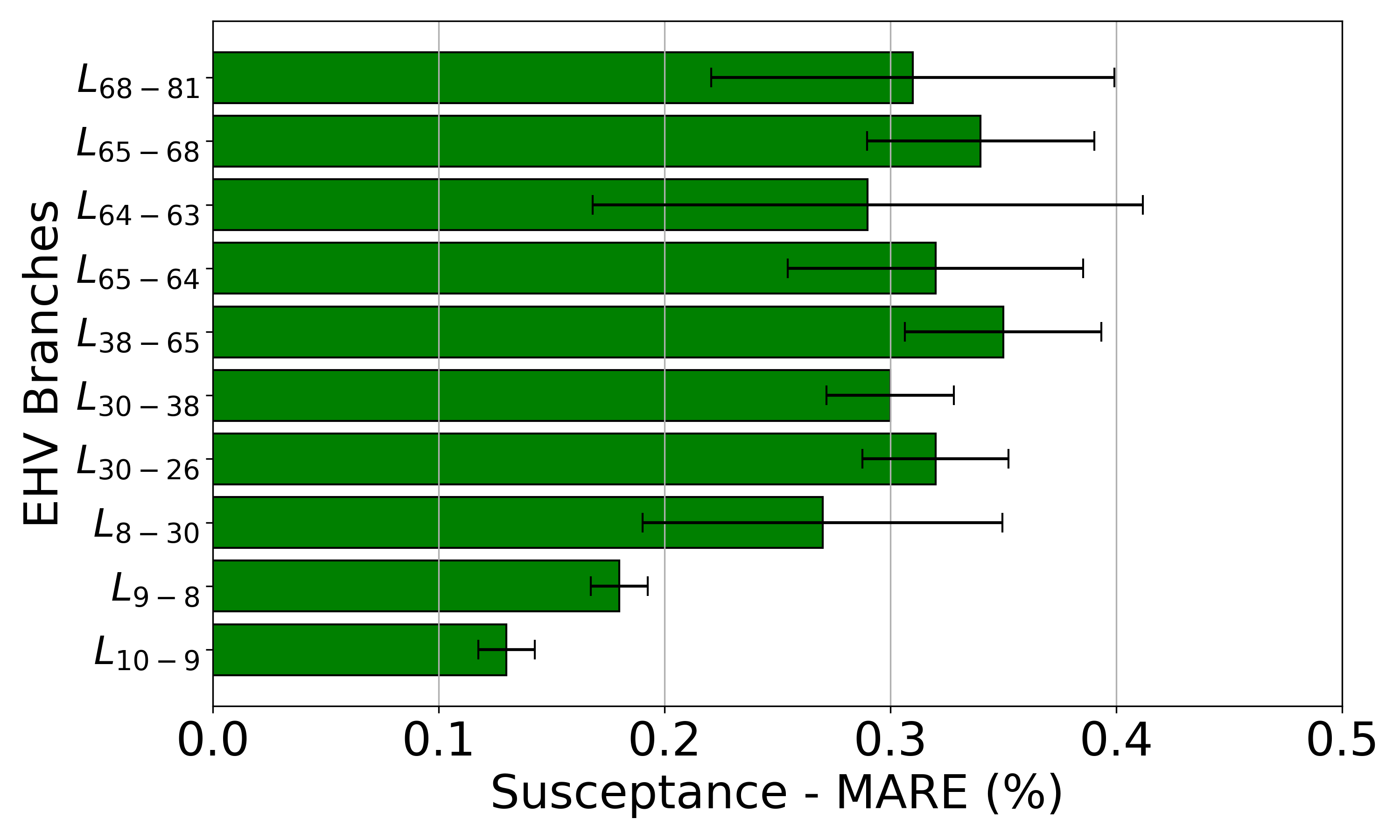} 
        \caption{Susceptance }
        \label{C3_Susceptance}
    \end{subfigure}
    \caption{Evaluation of line parameter estimates in presence of a non-ideal RQM pair and additive PMU noise}
\label{SLIC_Realistic_Evaluation_of_Line_parameters}
\vspace{-1em}
\end{figure*}
 
\begin{figure}[ht]
            \centering
            \vspace{-1em}
        \includegraphics[width=0.485\textwidth]{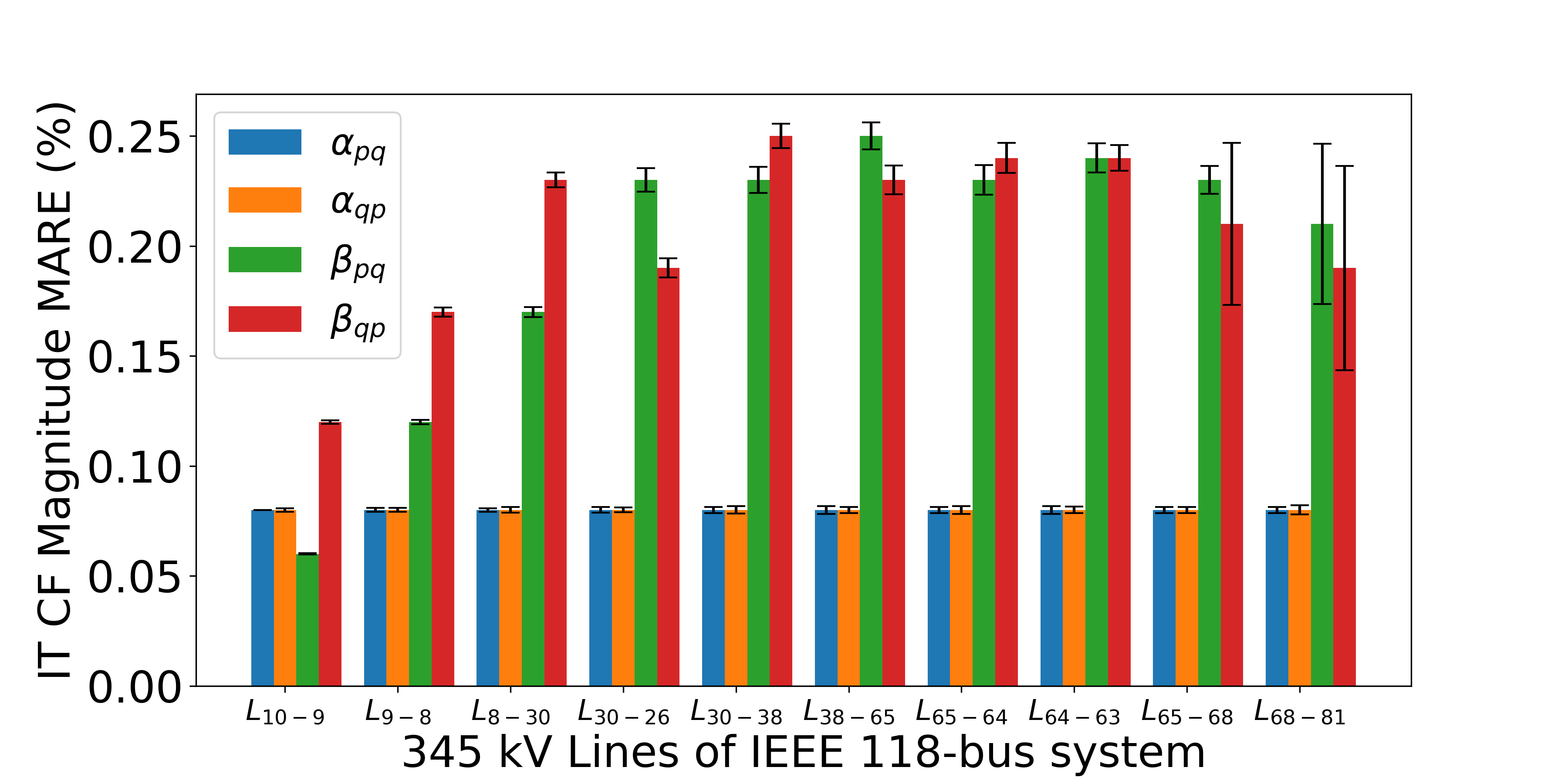}
            \caption{Accuracy of IT correction factor (CF) - magnitude, in presence of a non-ideal RQM pair and additive PMU noise}%
    \label{SLIC_Realistic_Evaluation_ITCF_Mag}
            \vspace{-1em}
\end{figure}

\begin{figure}[ht]
            \centering
\includegraphics[width=0.485\textwidth]{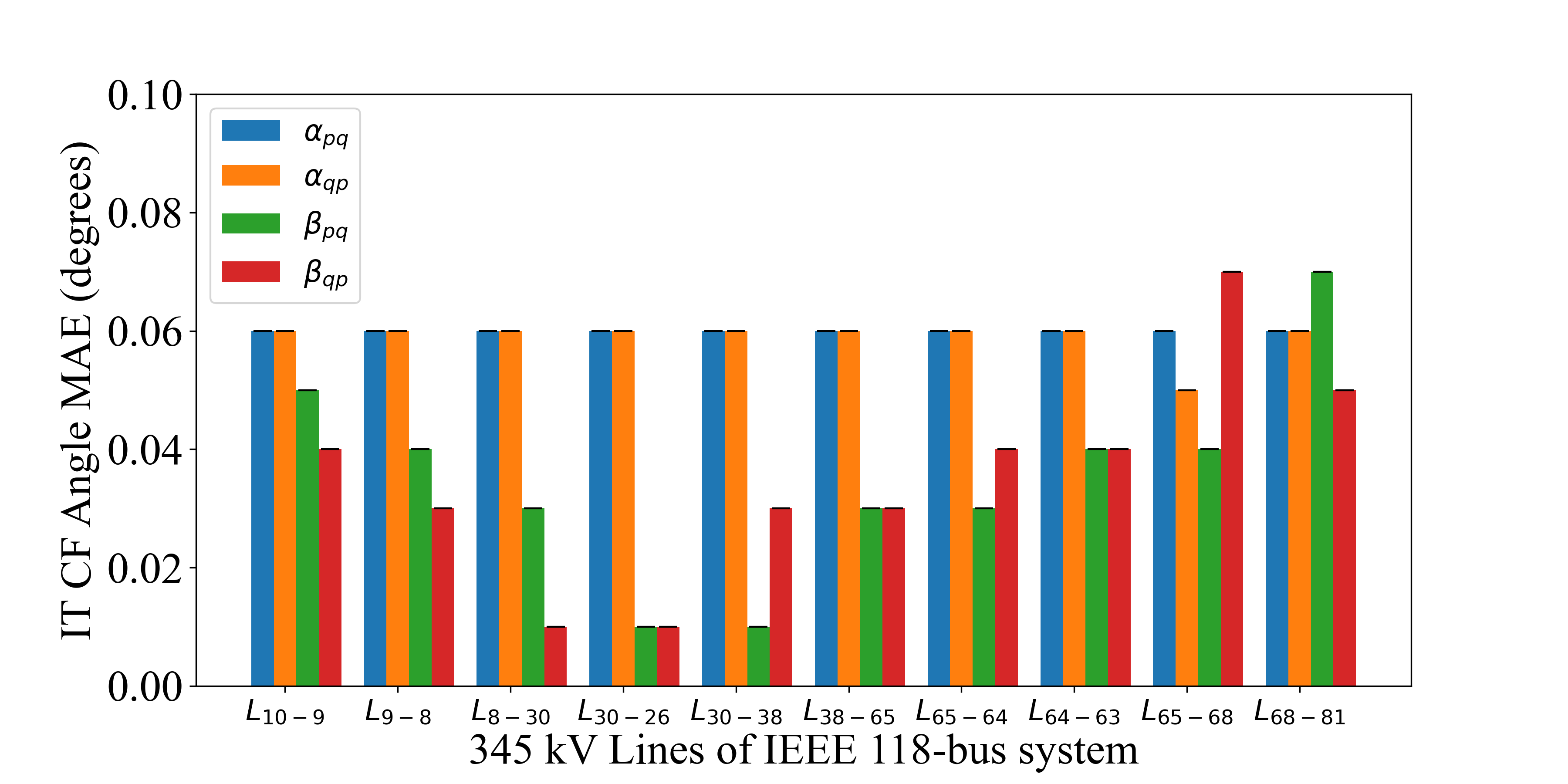}
            \caption{Accuracy of IT correction factor (CF) - angle, in presence of a non-ideal RQM pair and additive PMU noise }%
    \label{SLIC_Realistic_Evaluation_ITCF_Angle}
    \vspace{-1em}
\end{figure}

Finally, the most general scenario is considered in  which the $\eta$ of the RQMs lie within the bounds specified for the 0.15 IT accuracy class in \cite{IEEE_C57_13_2016_std_for_ITs}, 
and a noise of $0.1\%$ TVE is introduced
by the PMU.
The 
measurements generated based on this criteria are used as inputs to the NL-SLIC problem, which is solved using Algorithm \ref{alg:Sequential_NET_SLIC}.
The results for all $10$ lines are displayed in Figs. \ref{SLIC_Realistic_Evaluation_of_Line_parameters}-\ref{SLIC_Realistic_Evaluation_ITCF_Angle}.
Fig.  \ref{SLIC_Realistic_Evaluation_of_Line_parameters} demonstrates 
that the reactance and susceptance estimates are very accurate 
($\mathrm{MARE}<0.5\%$).
Even for the resistance estimates, the $\mathrm{MARE}$ is less than $3\%$ confirming the ability of the proposed approach to accurately estimate line parameters under realistic conditions.

Fig.~\ref{SLIC_Realistic_Evaluation_ITCF_Mag} and Fig.~\ref{SLIC_Realistic_Evaluation_ITCF_Angle} show the accuracy in estimating the IT correction factors for magnitudes and angles, respectively.
In Fig. \ref{SLIC_Realistic_Evaluation_ITCF_Mag} we see that even considering $\mathrm{MARE}$ and $\mathrm{SDARE}$, the errors in the magnitude estimates never exceeded $0.30\%$.
Similarly, from Fig. \ref{SLIC_Realistic_Evaluation_ITCF_Angle}, we see
that the maximum error in the angle estimates remained below $0.08^{\circ}$ for all the lines.
The consistently low $\mathrm{SDARE}$ indicated by the short height of the whiskers, particularly for the VT correction factor estimates in Figs. \ref{SLIC_Realistic_Evaluation_ITCF_Mag} and \ref{SLIC_Realistic_Evaluation_ITCF_Angle}, are a testament to the robustness of the proposed approach for SLIC.

\begin{figure}[ht]
\vspace{-0.5em}
            \centering
        \includegraphics[width=0.485\textwidth]{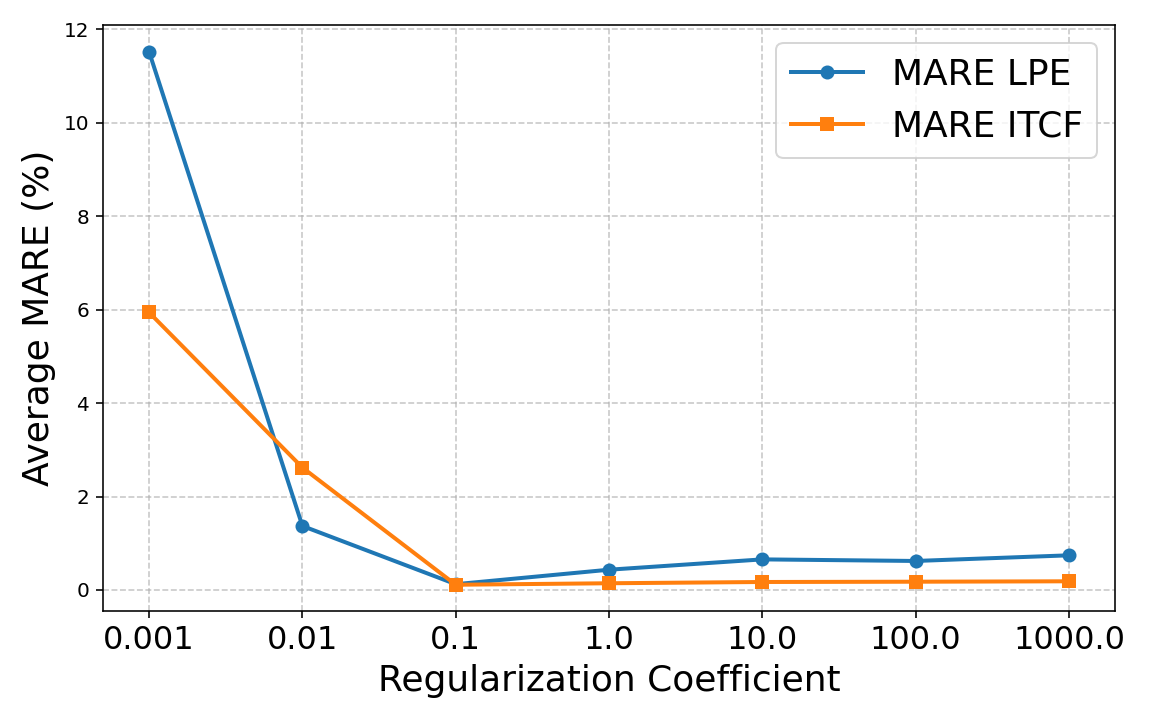}
        \vspace{-1.5em}
            \caption{\textcolor{black}{Sensitivity of  regularization coefficient}}
    \label{Lambda_sensitivity2}
    \vspace{-1em}
\end{figure}

\vspace{-1.25em}
\textcolor{black}{
\subsection{Sensitivity Analysis for Regularization Coefficient}
\label{Lambda_sensitivty}
A sensitivity analysis was conducted to analyze the impact of the value of the regularization coefficient (denoted by $\lambda$ in \eqref{D_SLIC_RQM_branch_with_Regularizer} and $\lambda_1$ in \eqref{Branch_Pair_Wise_Eq_constr_Optimization}). The value of the regularization coefficient was varied from $0.001$ to $1000$ on a log-scale while keeping everything else
the same. The experiment was repeated $100$ times for each value of the regularization coefficient 
and the resulting $\mathrm{MARE}$s for both line parameters as well as IT correction factors were recorded. The obtained error metrics are presented in Fig. \ref{Lambda_sensitivity2}.}

\textcolor{black}{From Fig. \ref{Lambda_sensitivity2} it can be noted that both the line parameters as well as the IT correction factors yield optimal results (i.e., lowest error index) when the value of the regularization coefficient is $0.1$.   
For cases in which the regularization term carries negligible weight relative to the primary objective function term (i.e., $\lambda$ and $\lambda_1$ much lower than $0.1$), the estimation error is significantly large. Conversely, as $\lambda$ and $\lambda_1$ increase beyond $0.1$, the error in line parameter estimation begins to rise gradually. 
Hence, it can be concluded that a suitable value for the regularization coefficient is $0.1$,
which is the value that was used in the simulations (see Section \ref{Simulated_Data_NETSLIC_results_IEEE118bus}).
}

\vspace{-0.5em}
\subsection{Comparison with a State-of-the-Art Approach}
\label{ComparisonSOTA}

\begin{figure*}
    \centering
    \begin{subfigure}[b]{0.67\columnwidth}
        \centering
        \includegraphics[width=\textwidth]{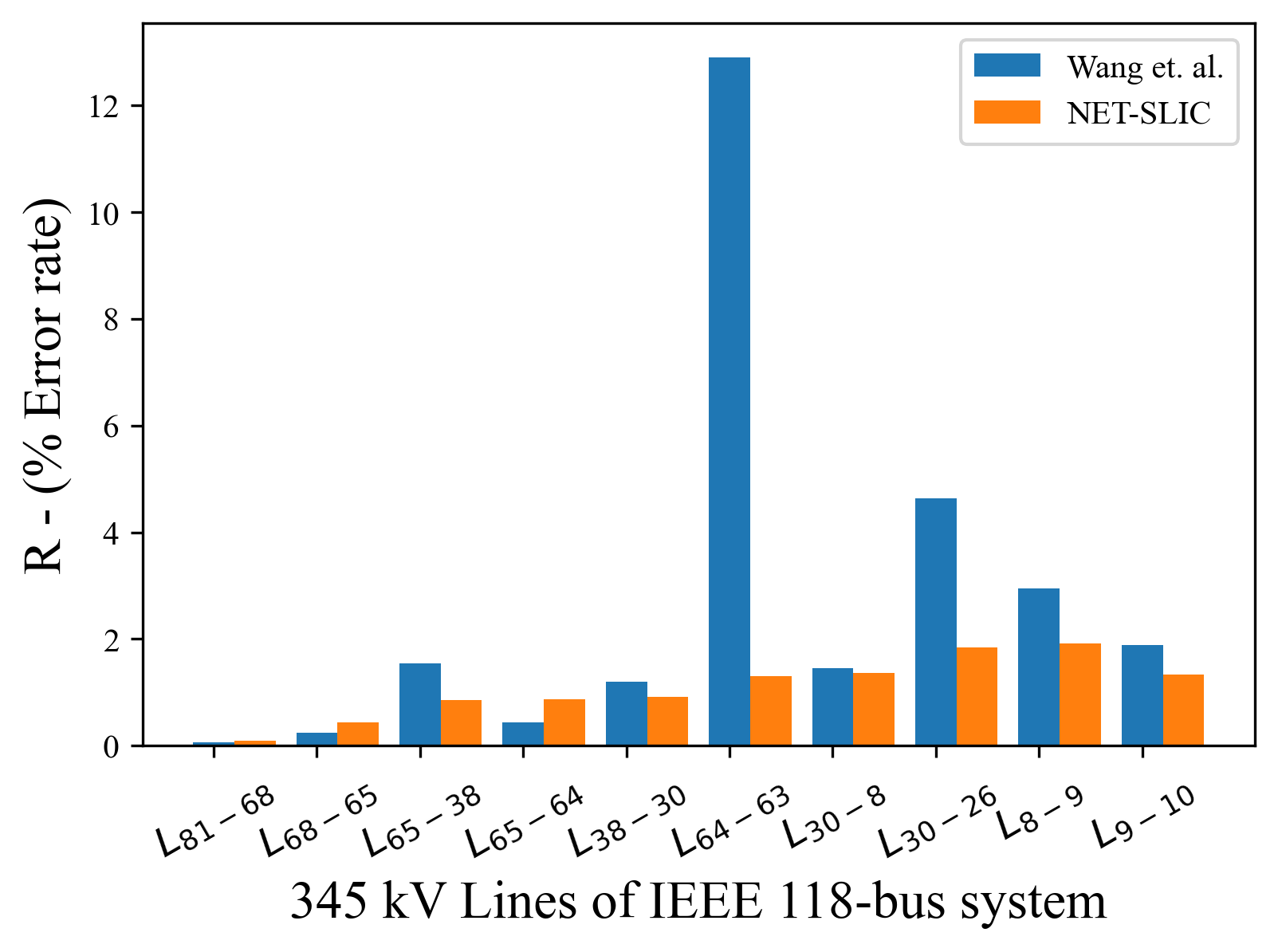}  
        \caption{Resistance}
        \label{R_ARE_Wang_NETSLIC_Comparison}
    \end{subfigure}
      \hfill
    \begin{subfigure}[b]{0.67\columnwidth}
        \centering
        \includegraphics[width=\textwidth]{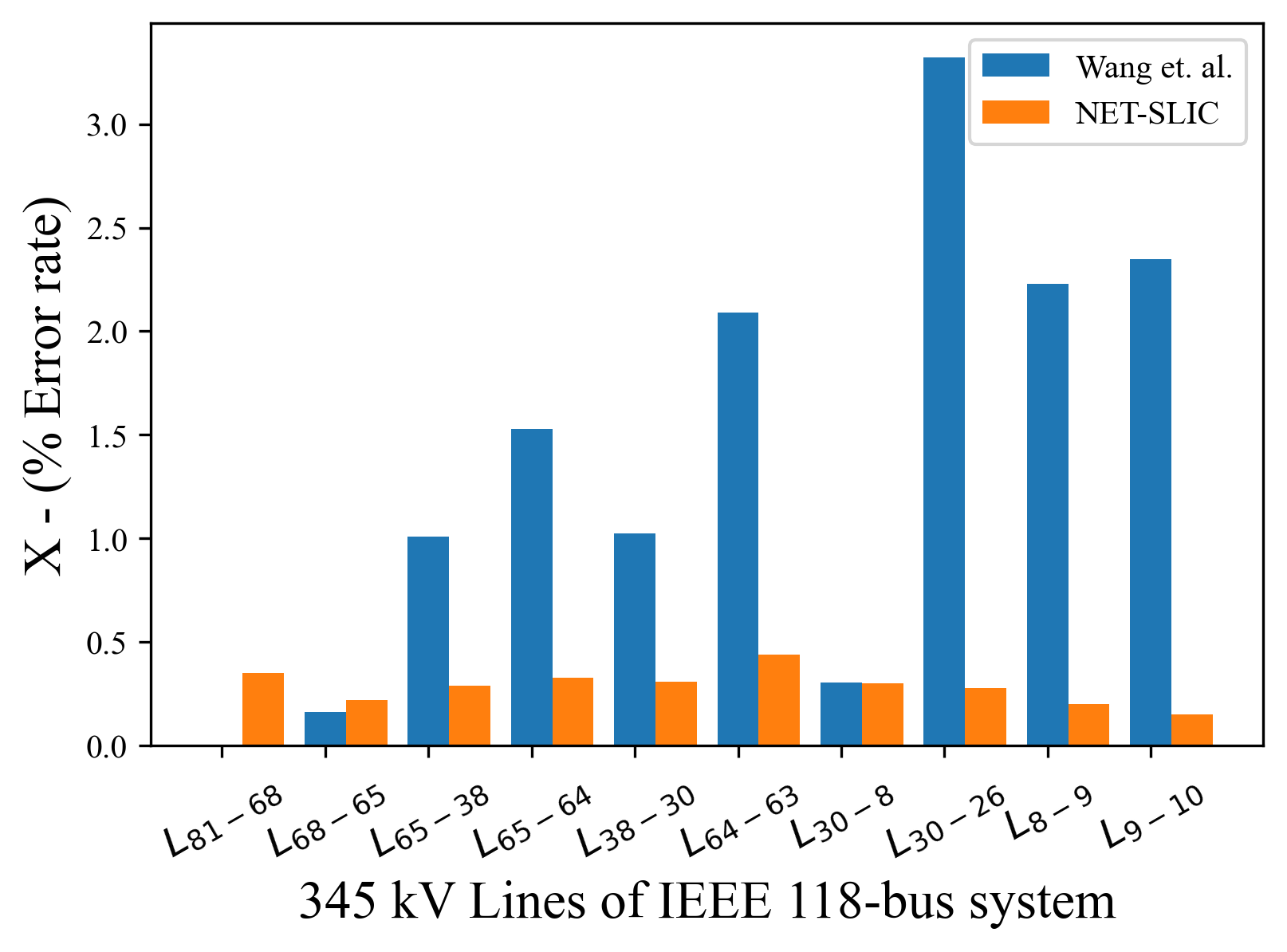}  
        \caption{Reactance}
        \label{X_ARE_Wang_NETSLIC_Comparison}
    \end{subfigure}
      \hfill
    \begin{subfigure}[b]{0.67\columnwidth}
        \centering
        \includegraphics[width=\textwidth]{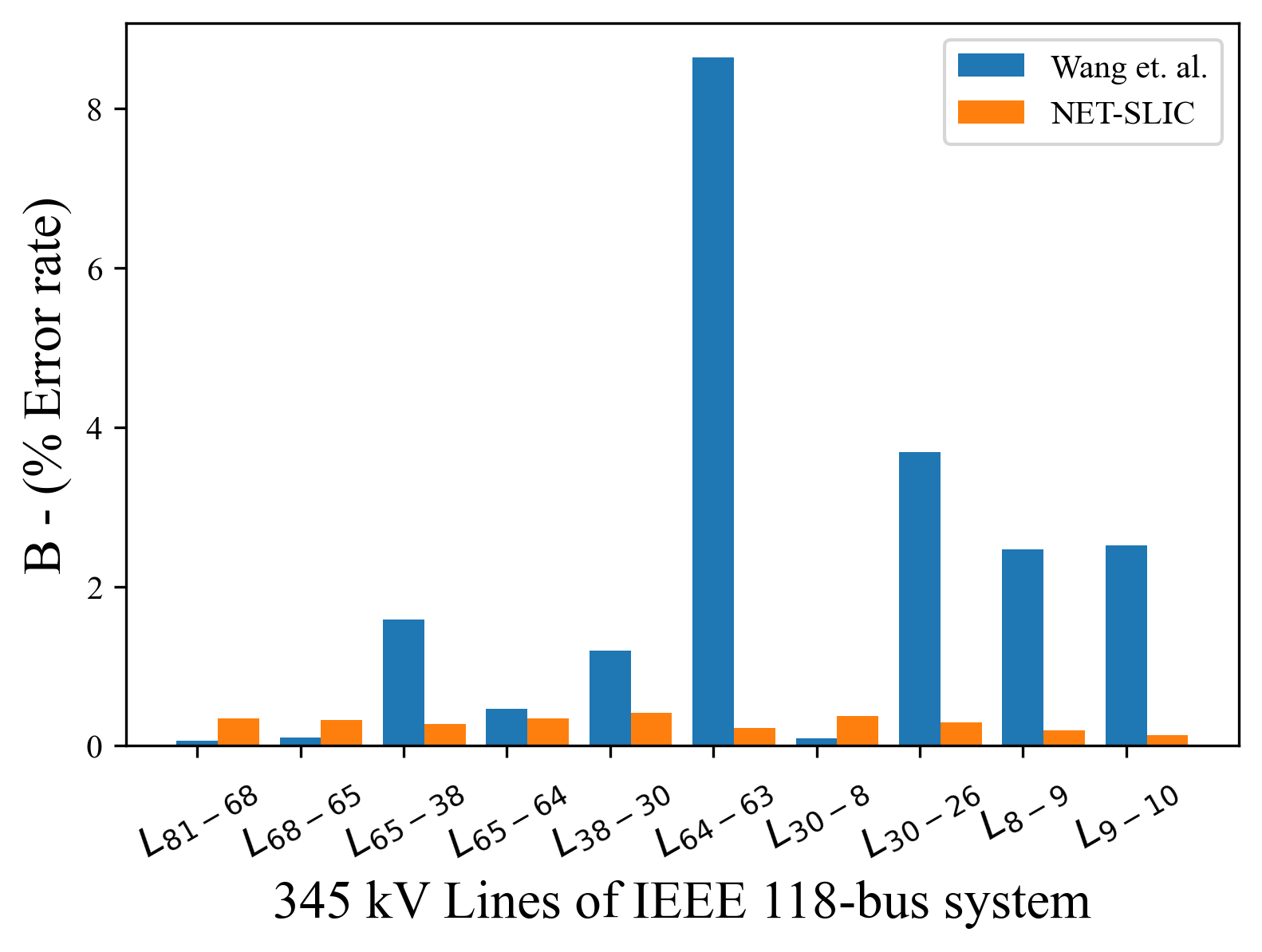} 
        \caption{Susceptance}
        \label{B_ARE_Wang_NETSLIC_Comparison}
    \end{subfigure}
    \caption{Comparison of LPE errors in $\%$ between the approach developed in Wang et al. \cite{wang2019transmission} and NET-SLIC}
    \label{RXB_ARE_Wang_NETSLIC_Comparison}
\end{figure*}

We compare the performance of the proposed approach (NET-SLIC)
with the approach developed by Wang et al. in \cite{wang2019transmission}.
Reference \cite{wang2019transmission}
was selected because it also performed SLIC for the 345 kV network of the IEEE 118-bus system, enabling a direct comparison.
To ensure a fair evaluation,
the PMU noise added to the phasor measurements
was made to match
the noise levels used in \cite{wang2019transmission}.
\textcolor{black}{Specifically, \cite{wang2019transmission} assumed a quantization scale of 12.16 V 
for the 345 kV voltage level, which corresponded to a maximum error magnitude of $3.52 \times 10^{-6}$ in p.u. Assuming a zero mean Gaussian distribution with this maximum error as the $3\sigma$ bound, the equivalent standard deviation became $1.17 \times 10^{-6}$  p.u.
This PMU noise characteristic was adopted for the simulation done in this sub-section.}
The accuracy of the line parameter and IT correction factor estimates obtained by the two methods is compared in Fig. \ref{RXB_ARE_Wang_NETSLIC_Comparison} and Tables \ref{Wang_etal_comparison_VTCF} and \ref{Wang_etal_comparison_CTCF}.

From 
Fig. \ref{R_ARE_Wang_NETSLIC_Comparison}-\ref{B_ARE_Wang_NETSLIC_Comparison},
it is evident that the proposed NET-SLIC significantly outperforms the method developed in
\cite{wang2019transmission} for LPE. 
In \cite{wang2019transmission}, the $\mathrm{ARE}$ goes as high as  $12.89\%$ in case of the resistance parameter and $8.64\%$ in case of the susceptance parameter, whereas the NET-SLIC results never crossed $2\%$ for any of the parameters.
It can be observed from Table \ref{Wang_etal_comparison_VTCF} that both the proposed method as well as the method developed in \cite{wang2019transmission} are extremely accurate in estimating the correction factors for VTs as evidenced by the very small 
absolute error values in real and imaginary coordinates (fourth decimal place or smaller). On the other hand, comparing the correction factors of CTs in Table \ref{Wang_etal_comparison_CTCF}, it can be noticed that the proposed method is often an order of magnitude more accurate (in the second or third decimal place) when compared to 
\cite{wang2019transmission}. Overall, it can be observed that the proposed method has either similar or better performance than \cite{wang2019transmission} for the estimation of line parameters and IT correction factors.

\begin{table}[ht]
\caption{VT correction factor comparison }
\label{Wang_etal_comparison_VTCF}
\begin{tabular}{|l|l|l|l|l|}
\hline
                &  \begin{tabular}[c]{@{}l@{}} $\mathrm{AE}$ in\\     $\mathcal{R}e \left(\alpha_{pq}\right)$ \end{tabular}   &
                \begin{tabular}[c]{@{}l@{}} $\mathrm{AE}$ in\\    $\mathcal{I}m \left(\alpha_{pq}\right)$ \end{tabular} 
                & \begin{tabular}[c]{@{}l@{}} $\mathrm{AE}$ in\\     $\mathcal{R}e \left(\alpha_{qp}\right)$ \end{tabular} & \begin{tabular}[c]{@{}l@{}} $\mathrm{AE}$ in\\    $\mathcal{I}m \left(\alpha_{qp}\right)$ \end{tabular}   \\ \hline
Wang et al. \cite{wang2019transmission}     & 0.00002    & 0.00004    & 0.00004    & 0.00004    \\ \hline
\begin{tabular}[c]{@{}l@{}} NET-SLIC \end{tabular}  & 0.0008     & 0.0009     & 0.0008     & 0.0009     \\ \hline
\end{tabular}
\vspace{-1em}
\end{table} 

\begin{table}[ht]
\caption{CT correction factor comparison }
\label{Wang_etal_comparison_CTCF}
\begin{tabular}{|l|l|l|l|l|}
\hline
                &  \begin{tabular}[c]{@{}l@{}} $\mathrm{AE}$ in\\     $\mathcal{R}e \left(\beta_{pq}\right)$ \end{tabular}   &
                \begin{tabular}[c]{@{}l@{}} $\mathrm{AE}$ in\\    $\mathcal{I}m \left(\beta_{pq}\right)$ \end{tabular} 
                & \begin{tabular}[c]{@{}l@{}} $\mathrm{AE}$ in\\     $\mathcal{R}e \left(\beta_{qp}\right)$ \end{tabular} & \begin{tabular}[c]{@{}l@{}} $\mathrm{AE}$ in\\    $\mathcal{I}m \left(\beta{qp}\right)$ \end{tabular}   \\ \hline
Wang et al. \cite{wang2019transmission}     & 0.0067    & 0.00025   & 0.0112    & 0.0226   \\ \hline
\begin{tabular}[c]{@{}l@{}} NET-SLIC \end{tabular}  & 0.0021    & 0.00005     & 0.0022     & 0.00023     \\ \hline
\end{tabular}
\vspace{-1em}
\end{table}

\subsection{NET-SLIC Results Using Real PMU Data}
\label{Field}
The results obtained on implementing NET-SLIC on actual PMU data obtained from 
a U.S. power utility located in the Eastern Interconnection, is described in this section.
\textcolor{black}{Positive-sequence voltage and current phasor measurements were obtained from
230\,kV substations of the partner utility. The raw data were first cleaned
to retain only steady-state operating intervals, removing any segments
containing transient events or disturbances. The original measurements had a
reporting rate of 30 frames per second; however, using them directly led to
ill-conditioning of the matrices in the estimation formulation. It was
observed that downsampling the data significantly improved the conditioning.
This is attributed to the fact that downsampled data retain measurements
from more distinct operating conditions, thereby reducing collinearity among
the data points. Based on this observation, the data were downsampled to one
frame per minute. The database values of the line parameters were used as the
initial guess for the LPE component, while 
ideal
values (i.e., unity real part and zero for imaginary part) were assumed as the initial
guess for the IT calibration component.}

A significant challenge in working with field data
is the absence of knowledge of ground truth (i.e., actual value of the parameters), making it difficult to evaluate performance directly. To overcome this challenge, we look at two aspects: \textit{consistency} of line parameter estimates and \textit{redundancy} of voltage measurements, as explained below. 

The \textit{consistency} in the line parameter estimates 
across different weekdays is realized from
Table \ref{OGnE_LPE_Data_1}.
It can be observed from the table that the parameters are generally consistent across all five days, particularly in the mornings (8 AM to 11 AM).
The higher variability in the afternoons (3 PM to 6 PM) could be attributed to increased fluctuations in temperature.
Overall, it can be concluded from the table that (a) line parameters do change over time, and (b) NET-SLIC is able to track the changes in a consistent manner.

\begin{table*}[ht]
\centering
\caption{\color{black}Line parameter estimates (in p.u.) obtained using NET-SLIC with real PMU data}
\vspace{-0.5em}
\label{OGnE_LPE_Data_1}
\color{black}
\begin{tabular}{|c|c|c|c|c|c|c|c|}
\hline
& & \multicolumn{2}{c|}{Resistance} & \multicolumn{2}{c|}{Reactance} & \multicolumn{2}{c|}{Susceptance} \\ \hline
& & 8 AM--11 AM & 3 PM--6 PM & 8 AM--11 AM & 3 PM--6 PM & 8 AM--11 AM & 3 PM--6 PM \\ \hline
\multirow{5}{*}{\rotatebox{90}{Estimated}} 
& Monday    & 0.0036 & 0.0037 & 0.0262 & 0.0262 & 0.0046 & 0.0045 \\ \cline{2-8}
& Tuesday   & 0.0036 & 0.0041 & 0.0261 & 0.0259 & 0.0047 & 0.0046 \\ \cline{2-8}
& Wednesday & 0.0036 & 0.0042 & 0.0260 & 0.0259 & 0.0054 & 0.0054 \\ \cline{2-8}
& Thursday  & 0.0037 & 0.0045 & 0.0259 & 0.0258 & 0.0053 & 0.0056 \\ \cline{2-8}
& Friday    & 0.0037 & 0.0048 & 0.0257 & 0.0254 & 0.0059 & 0.0061 \\ \hline
\multicolumn{2}{|c|}{Database value} & \multicolumn{2}{c|}{0.0044} & \multicolumn{2}{c|}{0.025} & \multicolumn{2}{c|}{0.0052} \\ \hline
\end{tabular}
\end{table*}

\begin{figure}
            \centering
            \vspace{-2em}
        \includegraphics[width=0.485\textwidth]{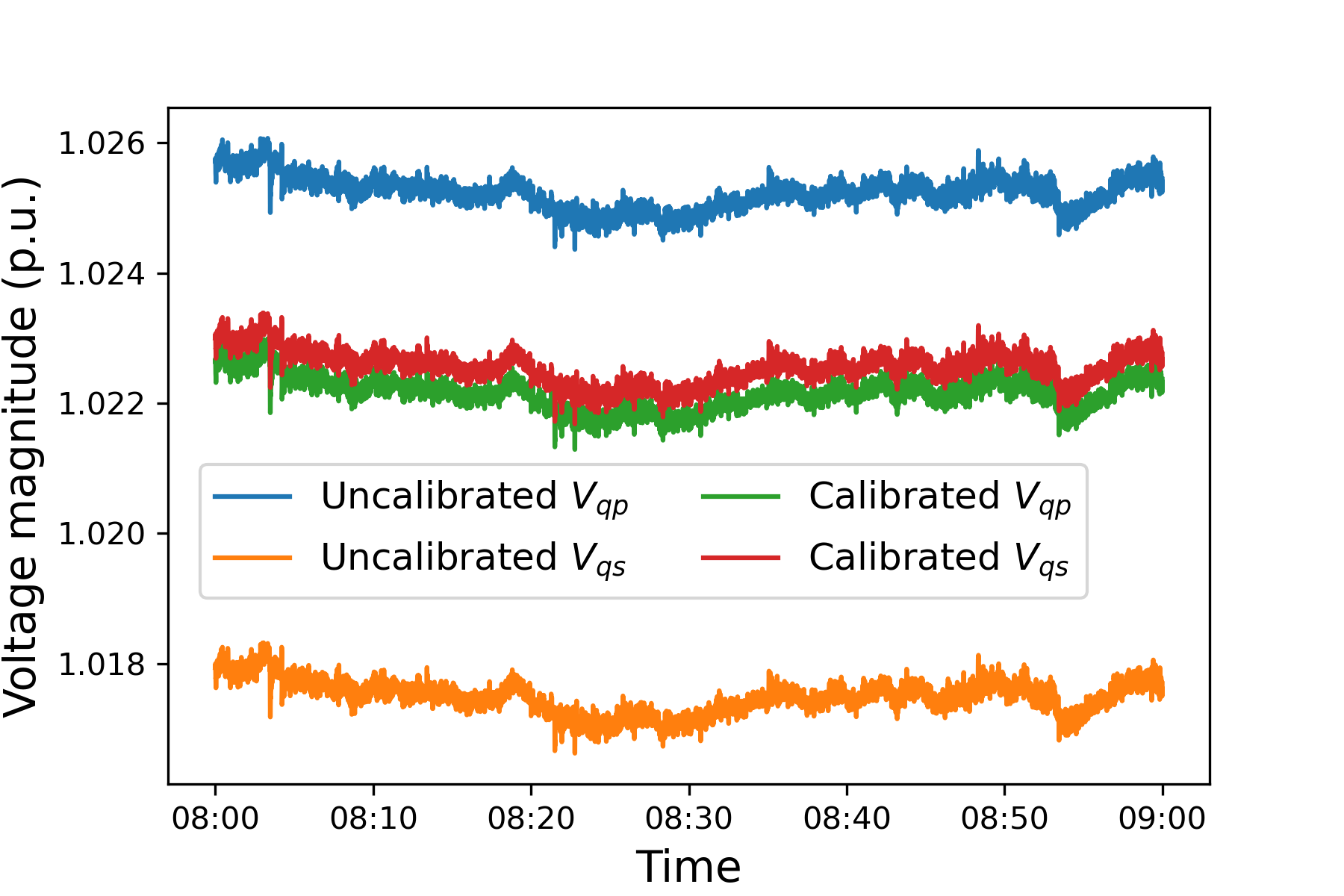}
            \vspace{-1.5em}
            \caption{Voltage magnitude measurements before and after calibration}%
    \label{SLIC_OGnE_Redundancy_Vmag}
\end{figure}

\begin{figure}
            \centering
            \vspace{-1.5em}
        \includegraphics[width=0.485\textwidth]{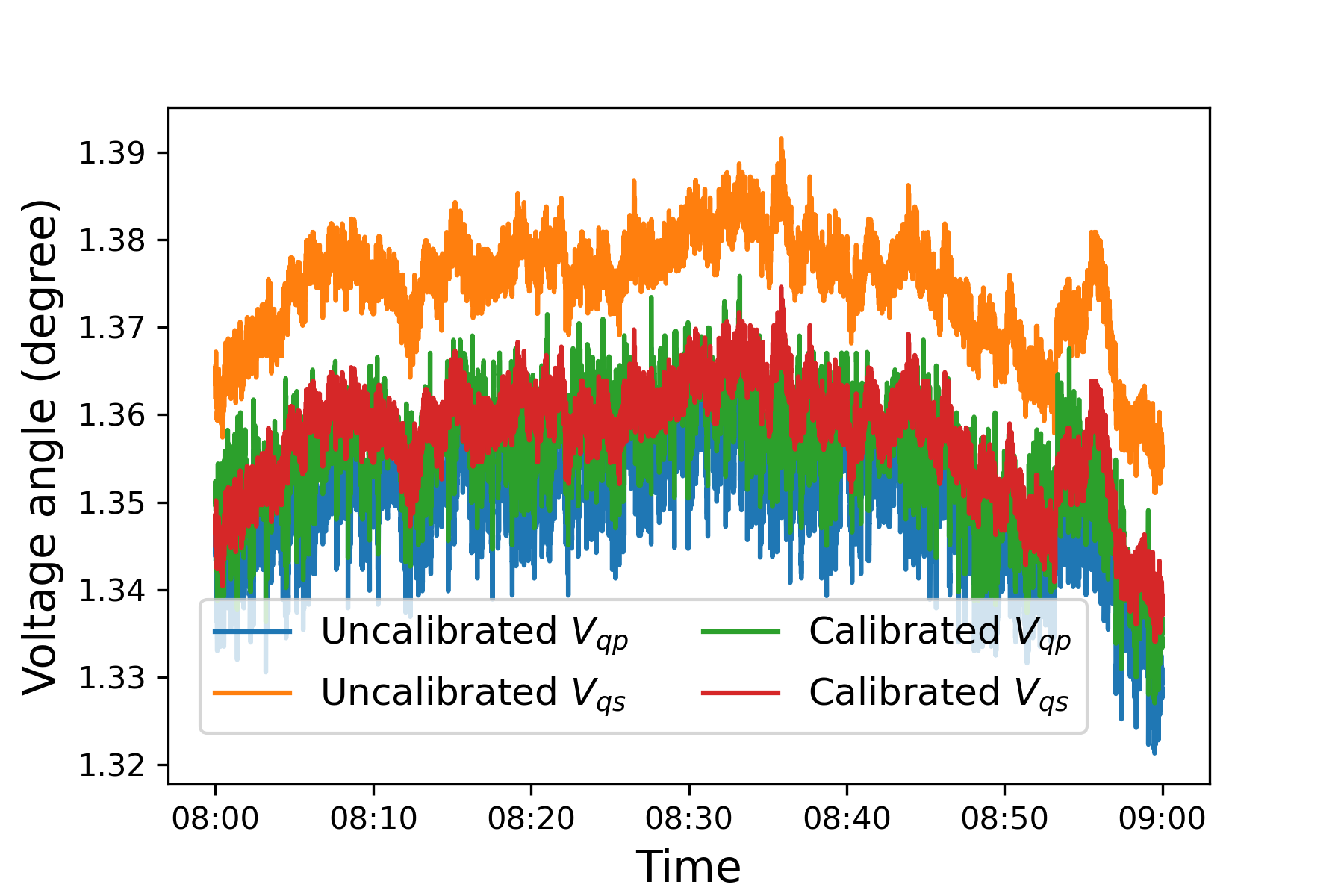}
            \vspace{-1.5em}
            \caption{Voltage angle measurements before and after calibration}
    \label{SLIC_OGnE_Redundancy_Vang}
            \vspace{-1em}
\end{figure}

To overcome the lack of knowledge of ground truth for the IT correction factors, we leverage the \textit{voltage redundancy property}, which involves observing the voltage of the same bus by two different PMUs (see Section \ref{VoltRe}).
To illustrate, consider the uncalibrated voltage magnitudes of $V_{qp}$ and $V_{qs}$\footnote{Here, $p$, $q$, and $s$ are three consecutive buses of a connected tree present in the power system network of 
the U.S. power utility.} shown in \textit{blue} and \textit{orange} in Fig. \ref{SLIC_OGnE_Redundancy_Vmag}. Both PMUs are expected to observe the same voltage ($V_q^*$), but as seen in the figure, a \textit{scaled} difference exists between them. This discrepancy (scaled instead of random addition)
suggests that the cause is not PMU noise but ratio errors in the VTs for $V_{qp}$ and $V_{qs}$, respectively. A similar observation is made for the voltage angles, as illustrated in Fig. \ref{SLIC_OGnE_Redundancy_Vang}.
The calibrated voltage magnitudes and angles obtained through NET-SLIC are shown in Fig. \ref{SLIC_OGnE_Redundancy_Vmag} and Fig. \ref{SLIC_OGnE_Redundancy_Vang}, respectively, using \textit{green} and \textit{red} colors.
It is clear from the figures that the green and red plots are closer to each other than the blue and orange plots, indicating that the estimates of the same voltage from two different sensing systems have become similar after calibration.
In summary, from Table \ref{OGnE_LPE_Data_1} and Figs. \ref{SLIC_OGnE_Redundancy_Vmag} and \ref{SLIC_OGnE_Redundancy_Vang}, it can be inferred that NET-SLIC can reliably estimate line parameters and calibrate ITs for 
actual power systems.

\vspace{-1em}
\subsection{A Downstream Application of SLIC - Improvements in LSE}
\label{LSE}

Although a variety of power system applications are benefited from SLIC, we choose the LSE application here because it makes use of both 
line parameters as well as IT correction factors.
Consider a scenario where the observation matrix of LSE is created using the legacy line parameters from the power utility's database, implying that the actual values of the line parameters are different.
Next, assume that IT calibration has not been performed, i.e. the ITs associated with PMUs have ratio errors whose values are specified in 
\cite{IEEE_C57_13_2016_std_for_ITs}. 
This setup serves as the base-case for the LSE comparison study. 

To emulate this base-case in simulation, we again considered the IEEE 118-bus system.
We placed PMUs in this system in accordance with the algorithm developed in \cite{pal2014pmu} which not only ensures full observability but also places PMUs on the buses of choice  (all the 345 kV buses).
The net $\mathrm{ARE}$ and net $\mathrm{AE}$ across all the buses for the base-case LSE (i.e., without performing SLIC) was $1.34\%$ in magnitude and $0.97^{\circ}$ in angle, respectively.
Next, we perform SLIC only for the 345 kV network of this system.
The net $\mathrm{ARE}$ and net $\mathrm{AE}$ for LSE after performing SLIC decreased to $1.03\%$ in magnitude and $0.61^{\circ}$ in angle, respectively. This indicates an average increase in accuracy of $24\%$ in magnitude and $37\%$ in angle.

\begin{figure}[ht]
    \vspace{-0.5em}
    \centering
    \begin{subfigure}[b]{0.485\columnwidth}  
        \centering
         \includegraphics[width=\textwidth]{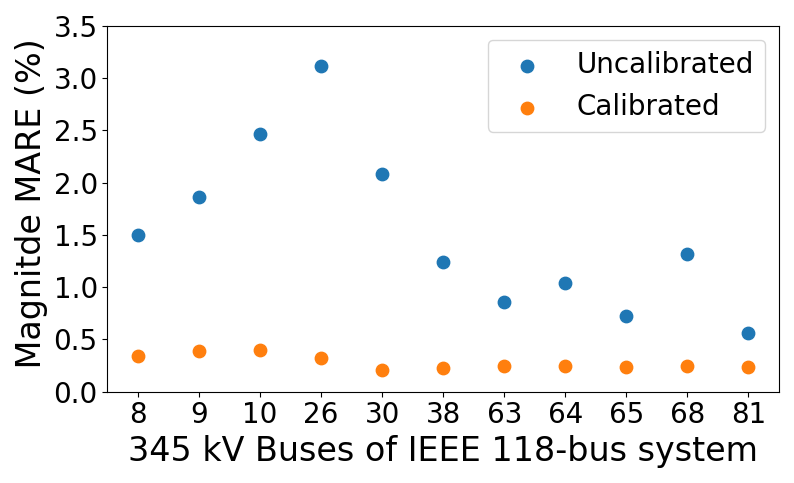} 
        \vspace{-1.5em} 
        \caption{Magnitude $\mathrm{ARE}$}
        \label{LSE - SLIC impact - 118 bus - Mag ARE}
    \end{subfigure}
    \hfill
    \begin{subfigure}[b]{0.485\columnwidth}  
        \centering
        \includegraphics[width=\textwidth]{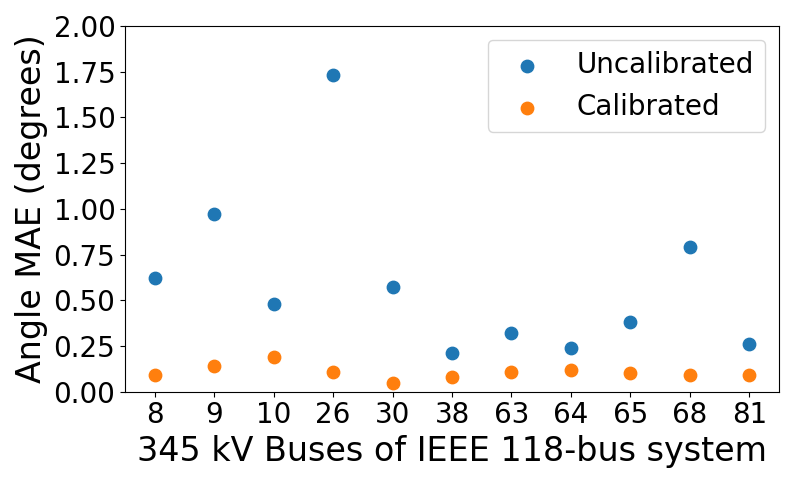} 
        \vspace{-1.5em}
        \caption{Angle $\mathrm{AE}$}
        \label{LSE - SLIC impact - 118 bus - Angle AE}
    \end{subfigure}
    \caption{Improvements in estimating states of the 345 kV network of IEEE 118-bus system because of SLIC}
    \label{LSE - SLIC impact - 118 bus}
    \vspace{-1em}
\end{figure}

The biggest improvements in the state estimation accuracy occurred in the 345 kV network; see Fig. \ref{LSE - SLIC impact - 118 bus}.
From this figure, 
it can be realized that performing SLIC helped in bringing the magnitude state estimate errors in $\mathrm{ARE}$ from high values such as $2.3\%$ in bus 10 and $1.8\%$ in bus 30 to values well below $0.4\%$.
Similarly, in the case of angles, relatively higher $\mathrm{AE}$ in angle estimates are brought well below $0.25^{\circ}$ by performing SLIC
as depicted in Fig.  \ref{LSE - SLIC impact - 118 bus - Angle AE}.

\section{Conclusion}
\label{D_SLIC_Conclusions}
The interdependent nature of LPE and IT calibration necessitates the creation of an approach that can simultaneously perform both.
This is the basis of the SLIC problem, and a novel methodology to solve this problem without making unrealistic assumptions is developed in this paper.
The proposed solution employs power system domain knowledge
as regularization terms and equality constraints, which are then incorporated into an optimization formulation. 
To  solve this formulation, a Newton's method enhanced with a trust-region approach, called NET-SLIC, has been employed.
The effectiveness of NET-SLIC in solving the SLIC problem is 
demonstrated using both simulated as well as actual PMU data. Finally, the potential for significant improvements in downstream applications, such as LSE, was illustrated, which showcases the practical value of performing SLIC.

\bibliographystyle{IEEEtran}
\bibliography{ References/D_SLIC_References}

\end{document}